\documentclass[preprint,review,12pt] {elsarticle}

\usepackage{amssymb}
\usepackage{amsmath}
\usepackage{verbatim}
\usepackage{dcolumn}
\usepackage{epsfig}

\begin{document}
\overfullrule=1pt
\hbadness=6000
\hfuzz=2pt
\begin{frontmatter}



\title{Nuclear Equation of State and many-body phase-space correlations 
in the  
Constrained Molecular Dynamics}


\author{M.Papa}

\affiliation[aff1]{organization={INFN-Catania},
            addressline={Via S.Sofia 64}, 
            city={Catania},
            postcode={95123}, 
            state={Italy},
            country={Sicily}}

\begin{abstract}
Many-body correlations  characterizing the Constrained  Molecular Dynamics (CoMD)are analyzed in the case of finite and zero range effective microscopic interactions. The study begins by analyzing  the  case of infinite nuclear matter at zero temperature. 
A comparison with the predictions in the mean-field
(MF) limit corresponding to different effective masses, highlights non-negligible differences regarding the produced Equation of State (EoS).
A procedure is illustrated  to determine the necessary corrections of the effective interaction parameters in the CoMD model so to reproduce the chosen EoS. 
The specific model calculations,
the general feature of the discussed correlations gives a wider meaning to the resulting differences, which are in fact strongly related both to the Pauli principle constraint and to the localization effects related to wave-packets dynamics.
Moving on to finite systems,  preliminary results are shown in relation to the reaction mechanisms in the $^{64}Ni+^{48}Ca$  system  described by the CoMD model. 
Therefore, the topic covered illustrates the effects produced by the correlations of the CoMD dynamics on the EoS  and on some observables commonly studied  in heavy ion collisions.  

\end{abstract}




\end{frontmatter}


\section{1.Introduction}
The description of many-body systems is one of the most
difficult problems in nuclear physics due to their complexity being quantum objects described by
a large number of degrees of freedom. A large variety of theoretical
models have been developed using mean-field (MF)
and beyond-mean-field  based approaches 
\cite{dft0,dft,neg,brink,tra2022}. 
In these
approaches, which use the independent particle approximation
as a starting point, phenomenological effective interaction like
Skyrme and Gogny forces are widely used, taking advantage
of their simple form \cite{skyrmes,gogny,hfbp,hfb}.
Each of them produces energy density functional (EDF)
with parameters adjusted to reproduce the basic nuclear bulk
properties in the valley of stability and other properties related
to finite systems such as binding energies and nuclear radii.
Heavy ion collisions (HIC) at energies well above 
the mutual Coulomb barriers are usually described through semi-classical
approaches  based on MF, BUU models or 
quantum molecular dynamics approaches (QMD) \cite{buss,onorev,aich,tra2016,tra2018,fiss,comd}. These last describe the single
particle wave functions by means of well localized 
wave-packets (WPs) with fixed widths.  
In this way many-body correlations  are produced which  lead to the spontaneous formation of clusters.
In these semi-classical approaches the effective interaction 
obviously plays a key role, and in many cases it just represents the main subject of investigation. 
The class of Skyrme microscopic zero-range (ZR) interactions \cite{skyrmes} generally produce 
a simple density functional. 
In the MF framework 
momentum-dependent effects are introduced through a short-range expansion of a finite-range interaction. 
It produces a momentum dependent interaction (MDI).  
In the Gogny interaction \cite{gogny}, the finite range 
is explicitly introduced  through  Gaussian terms depending on the two-nucleons distance and on the related widths which define the interaction ranges. 
Additional terms are
also considered: a density dependent ZR interaction and a spin-orbit interaction.
Many parameters are necessary 
to reproduce a large variety of nuclear  properties in infinite and finite systems.
A more practical form has been proposed  
for MF transport models applied to 
heavy-ion collisions 
\cite{mdi}.
The non-locality of the effective interactions produces 
corrective factors to the in-medium nucleon kinetic energy formally represented through a density dependent nucleon effective mass $m^{*}$. 
Different effective masses for neutrons and protons mainly due to the iso-vectorial interaction 
are also able to affect several astrophysical processes and the structure 
of neutron-rich nuclei \cite{effmass,massnp}.
\noindent

In several cases MF and molecular dynamics approaches  share the same  microscopic effective interactions \cite{tra2022}. 
From a general point of view, it can be  expected
that the typical and explicit two or many-body correlations of QMD-like 
approaches could instead play a  role in many-body quantities such as the total energy.
In these cases, therefore, it would be desirable to investigate  
at what extent these specific correlations may affect the functional related to the total energy. 
Consequently  one should establish the possible related corrections with respect to the commonly used
parameter values. 
In reference \cite{miosky} this problem was investigated   
by using the CoMD model\cite{comd}. 
The COMD model belongs to the wide class of QMD-like models.
Its peculiarity is that the equations of motion associated to the centroids of the wave packets are constrained (by numerical procedures) in such a way to produce phase-space distribution which locally satisfy the Pauli principle at every time step.

In Ref. \cite{miosky} a ZR
Skyrme interaction was used, and the calculations were performed for a large sphere
of nuclear matter (NM). Moreover, in that study  were included all the  main processes
(warming-cooling procedure for the energy minimization coupled with the Pauli principle constraint, selection of the most "stable" configurations) which are used to describe finite nuclei in 
long-lived configurations.
In this work we want to extend these kinds of studies around the saturation density and beyond it to the case of
a finite range interaction, and in the case of  NM  we limit to  zero temperature. 
Moreover for both cases MDI and ZR we want
to evaluate possible consequences of these correlations on some measurable quantities obtained from the study of   an Heavy Ion Collisions (HIC). 
\noindent
In particular,
in Sect. 2.1 the used effective microscopic interaction  is specified.
In Sect. 2.2 we evaluate it in the mean-field limit trying to reproduce the main
properties of NM at the saturation density.

In Sect. 2.3 are specified the corresponding expressions  obtained in CoMD model 
and Sect. 2.4 is dedicated to a general  discussion about the source of many-body correlations.
In this work NM is simulated trough 
box calculations with periodic boundary
conditions.
In Sect. 2.5 the comparison between the model calculations
and the MF predictions is analyzed and  the necessary corrections on the parameters  are computed.
\noindent 
Details on the adopted constraint procedure and on the effects of the CoMD correlations on the occupation numbers in "ground state" configurations (numerical minimization of the total energy)  are given in  Appendix A.
In  Appendix B the results of the above comparisons in the
case in which a simpler ZR range interaction is used are discussed.
Appendix C shows the effects on the light cluster production at low density.
In Sec.3 proceeding to finite systems the  results of calculation concerning
the $^{64}Ni+^{48}Ca$ collision at different energies are shown. 
In particular, for the case of finite range
and ZR interaction is discussed the comparison of different reaction mechanisms and nucleon transverse flow production  for central, and semi-central collisions.

\noindent
Furthermore  to understand at what extent the correlations produced by the model can affect the above observables, comparisons using the set of  parameter values obtained in the MF scheme are also shown.
Summary and final remarks will be illustrated in Sec. 4.

\section{\label{sec:level1}The EDF}
The present study was conceived by referring to
certain functionals of the density related to the total and symmetry energies in NM at zero temperature. In fact these functionals are usually
 taken as a reference in the studies using the wide class of semi-classical transport models \cite{tra2022}. To test the different effective interactions, these approaches are used as a bridge to connect the information collected from heavy-ion collisions to the
 NM  equation of state (EoS)(see as some examples Refs. \cite{dip,lefevre,plbayes}.

\noindent
Therefore, in the following we will study  an example of EDF characterized by some properties at zero temperature that we choose as reference: equilibrium density 
$\rho_{0}=0.165$ $fm^{-3}$, associated 
binding energy
$E(\rho_{0})=-16$ MeV,  compressibility $K(\rho_{0})=240$ MeV,  
symmetry energy $E_{sym}=30$ MeV, an effective pairing energy 
$E_{\pi}=E_{S}- E^{q,ex}_{2}- E^{q,ex}_{3}- E^{q,ex}_{4}$ 
equal to -2 MeV per nucleon at the saturation density in finite systems (around mass 100).
Moreover, the functional will correspond in the MF limit to a
relative isoscalar effective  mass $m^{*}=0.67$, 
and  neutron-proton
effective mass splitting $m_{n}^{*}-m_{p}=0.4\beta$ \cite{effmass}.
 Other conditions have been chosen corresponding to the vanishing of the total energy at zero density and no negative curvature  in the  density range $2 \div 3\rho_{0}$.  
Finally, different values of the slope parameter associated to the
symmetry energy $L=3\rho_{0}(\frac{dE_{sym}}{d\rho})_{\rho_{0}}$
have been considered changing in the range $ 55 \div 105$ MeV as suggested from different investigations 
\cite{effmass,prcconstr,dip,plbayes,prex2}.

\noindent
We now proceed in fixing the structure of the effective microscopic interaction from which the energy functional of the density can be obtained according to the chosen models.
\subsection{The microscopic interaction.}
\noindent
The microscopic effective interaction in the original formulation of the CoMD model \cite{comd}  was a simple  ZR interaction of the Skyrme type. We used a 2-body plus a 3-body interaction.  A third term
 describes the iso-vectorial interaction. This form was widely used in QMD-like models
 and BNV-BUU approaches (see as an example Ref. \cite{bon}). 
 A comparable degree of simplicity is maintained in this study just to point-out  as clearly as possible 
the effects related to the finite range interaction in MF and QMD-like approaches. 
 The total microscopic interaction will be the sum
 of the following contributions:

\begin{align}
 V(\textbf{r},\textbf{r'})&= [P_{2}+2P_{3}(\frac{\rho}{\rho_{0}})^{\sigma-1}]\textit{e}^{-(\textbf{r}-\textbf{r'})^{2}/\mu^{2}}\\
V_{0}(\textbf{r},\textbf{r'})&= \frac{1}{\rho_{0}}[P_{20}+\frac{2P_{30}}{\sigma+1} 
(\frac{\rho}{\rho_{0}})^{(\sigma-1)}]\delta(\textbf{r}-\textbf{r'})
\end{align}
and
\begin{align}
V^{sy}(\textbf{r},\textbf{r'})&=[P_{4}
(\frac{\rho}{\rho_{0}})^{(\gamma-1)}
(2\delta_{\tau-\tau'}-1)]
\textit{e}^{-(\textbf{r}-\textbf{r'})^{2}/\mu^{2}}
\\
V_{0}^{sy}(\textbf{r},\textbf{r'})&=\frac{1}{\rho_{0}}P_{40}(\frac{\rho}{\rho_{0}})^{(\gamma-1)}(2\delta_{\tau-\tau'}-1)\delta(\textbf{r}-\textbf{r'})\\
V_{S}(\textbf{r},\textbf{r'})&=\frac{1}{\rho_{0}}P_{\pi}(\frac{\rho}{\rho_{0}})^{(\gamma-1)}\delta_{s+s'}\delta_{\tau-\tau'}\delta(\textbf{r}-\textbf{r'})
\end{align}   

$\tau$ and $s$ indicates the third components of the nucleon iso-spin and spin quantum numbers 
 respectively.
The  contributions in Eq.(1) and Eq.(3) represent a 
generalization of the terms reported in \cite{miosky} associated 
to the two, three-body and iso-vectorial interactions.
In analogy with the Gogny interaction Ref. \cite{gogny} , we have substituted the delta functions in 
the spatial relative coordinates with a Gaussian whose width defines the only range $\mu=1.1$ fm that is used in this representation.

\noindent
Always in analogy to Ref. \cite{gogny} and after different attempts to satisfy the different requests of the reference EDF,
it has been necessary to add
residual ZR terms associated to the two, three-body and iso-vectorial contributions as shown in Eq.(2) and Eq.(4).
\noindent  
Finally Eq.(5) represents a ZR spin-spin interaction inspired from references \cite{chamel,bur}. This contribution
is necessary to reproduce  an effective "pairing" energy (see next section) of about -2 MeV
at the ground state for finite system with mass around 100. At the same time, this further contribution is able to locally produce
in box calculations (see next section) small values of average total  spin at the stationary conditions \citep{brink}.
The introduction of this term produce a further contribution to the
symmetry energy.  For the sake of simplicity, in order to maintain the same functional density dependence of the symmetry
energy, the parameter $\gamma$ describing the density dependence of the spin-spin interaction is the same as the one associated to the iso-vectorial contribution.

\subsection{EDF in the MF limit.}
Keeping in mind the above mentioned EoS reference properties
it's now possible to evaluate the microscopic interaction in the MF
approximation.
By considering the limit of infinite NM  and using  plane-wave single particle wave functions in a very large volume V, the integration on the spatial coordinates can be easily obtained. 
Due to the Fermionic nature of the problem the two-body wave functions are the direct product of single particle wave functions for not-identical
particles and the anti-symmetrized one for couples of identical particles \citep{brink,mdi}.
This will give rise to a MDI interaction. 
In particular, for the two-body exchange contribution and for the generic couples the matrix element will be: 

\begin{equation}
\Delta E^{ex}_{2}(\textbf{k}_{i},\textbf{k}_{j})=-\dfrac{P_{2}}{V^{2}}<\phi_{\textbf{k}_{i}}(\textbf{r})\phi_{\textbf{k}_{j}}(\textbf{r'})|
\textit{e}^{-(\textbf{r}-\textbf{r'})^{2}/\mu^{2}}|\phi_{\textbf{k}_{i}}
(\textbf{r'})\phi_{\textbf{k}_{j}}(\textbf{r})>
\end{equation}
while for the direct contribution $dr$ 
\begin{equation}
\Delta E^{dr}_{2}(\textbf{k}_{i},\textbf{k}_{j})=
\dfrac{P_{2}}{V^{2}}<\phi_{\textbf{k}_{i}}(\textbf{r})\phi_{\textbf{k}_{j}}(\textbf{r'})|
\textit{e}^{-(\textbf{r}-\textbf{r'})^{2}/\mu^{2}}|\phi_{\textbf{k}_{i}}
(\textbf{r})\phi_{\textbf{k}_{j}}(\textbf{r'})>
\end{equation}

For the exchange term,
by folding the interaction  with plane waves we obtain:
\begin{equation}
\Delta E^{ex}_{2}(\textbf{k}_{i},\textbf{k}_{j})=-\dfrac{P_{2}}{V^{2}}\int_{V_{x}}\int_{V}\textit{e}^{i (\textbf{k}_{i}-\textbf{k}_{j} )\textbf{x}}
\textit{e}^{- \textbf{x}^{2}/\mu^{2}}d^{3}xd^{3}R
\end{equation}
with $\textbf{x}=\textbf{r}-\textbf{r'}$ and 
$\textbf{R}=\dfrac{ \textbf{r}+\textbf{r'} }{2}$.
The double integration on the spatial coordinates gives:
\begin{equation}
\Delta E^{ex}_{2}(\textbf{k}_{i},\textbf{k}_{j})=-\frac{P_{2}}{V}(\sqrt{\pi}\mu)^{3}\textit{e}^{-\mu^{2}(\textbf{k}_{i}-\textbf{k}_{j})^{2}/4}
\end{equation}
From Eqs. (8,9) we see that  
the integration on $d^{3}R$  gives a global contribution just equal  to $V$. This can be interpreted as the result of a kind of internal average associated with the usage of plane waves.

For the direct term we get:
\begin{equation}
\Delta E^{dr,id}_{2}(\textbf{k}_{i},\textbf{k}_{j})=
\int_{V_{x}}\int_{V}
\textit{e}^{- \textbf{x}^{2}/\mu^{2}}d^{3}xd^{3}R=\dfrac{(\sqrt{\pi}\mu)^{3}}{V}
\end{equation}
For the exchange term
a further integration on the single particle states  
results in the following contribution  to the potential energy per nucleon for protons
$p$ or neutrons $n$ ($q$) :
\begin{align}
 E^{q,ex}_{2}&=-\frac{9}{16}P_{2}\frac{\pi^{3/2}}{k_{F}^{6}}\rho I_{q}\equiv P_{2}R_{q}(\rho,\beta),
\\ 
I_{q}& =\int_{o}^{k_{Fq}}g(k)k^{2}dk   
\end{align}
with
\begin{eqnarray}
g(k)=\frac{2}{\mu k}\left[\textit{e}^{-\mu^{2}(k_{Fq}+k)^{2}/4}
-\textit{e}^{-\mu^{2}(k_{Fq}-k)^{2}/4}\right]\nonumber\\
+\sqrt{\pi} \left[erf(\mu(k_{Fq}+k)/2) +erf(\mu(k_{Fq}-k)/2)\right]        
\end{eqnarray}
$k_{Fq}$ and $k_{F }$represent the Fermi momentum for protons or
neutrons and the one for symmetric NM respectively.
 $\beta=\frac{\rho_{n}-\rho_{p}}{\rho}$ is the charge-mass asymmetry parameter.

\noindent 
The  total direct contribution per nucleon  
including also  the direct contribution of  
not-identical particles leads to:
\begin{eqnarray}
E^{dr}_{2}=\frac{1}{2}P_{2}(\sqrt{\pi}\mu)^{3}\rho
\end{eqnarray}

By proceeding in an analogous way,
the 3-body term will produce:
\begin{align}
E^{q,ex}_{3}&=\frac{2}{\sigma+1}P_{3}(\frac{\rho}{\rho_{0}})^{\sigma-1}R_{q}(\rho,\beta)\\
E^{dr}_{3}&=\frac{P_{3}}{\sigma+1}(\sqrt{\pi}\mu)^{3}
\frac{\rho^{\sigma}}{\rho_{0}^{\sigma-1}}
\end{align}

Finally the iso-vectorial contributions will results as
follows:
\begin{align}
\hskip -5pt
E^{q,ex}_{4}&=P_{4}(\dfrac{\rho}{\rho_{0}})^{\gamma-1}R_{q}(\rho,\beta);
\\
E^{dr}_{4}&=\dfrac{P_{4}}{2}(\sqrt{\pi}\mu)^{3}
\dfrac{\rho^{\gamma}}{\rho_{0}^{\gamma-1}}\beta^{2};
\end{align}

The contributions associated to the residual two and three-body ZR interaction
will produce:
\begin{equation}
E_{0}=\frac{1}{2}P_{20}\frac{\rho}{\rho_{0}}+\frac{1}{\sigma+1}P_{30}(\frac{\rho}{\rho_{0}})^{\sigma}
\end{equation}
the iso-vectorial contribution gives:
\begin{equation}
E_{0}^{sy}=\frac{1}{2}P_{40}(\frac{\rho}{\rho_{0}})^{\gamma}\beta^{2}
\end{equation}
The spin-spin contribution to the binding energy for zero total spin results in:
\begin{equation}
E_{S}=\frac{1}{8}P_{\pi}(\frac{\rho}{\rho_{0}})^{\gamma}(1+\beta^{2})
\end{equation}

Therefore the total energy per nucleon $E$ can be expressed as:
\begin{eqnarray}
E =E_{0}+E^{sy}_{0}+E_{S}+\sum_{q=n,p}\sum_{i=2,3,4}E^{q,ex}_{i}+\sum_{i=2,3,4}E^{dr}_{i}
+\sum_{q=n,p}T_{Fq}
 \end{eqnarray} 
$T_{Fq}$ is the kinetic energy contribution related to the Fermi motion. 
The single particle potential for symmetric matter due to the exchange term is: 
\begin{equation}
U=-\frac{1}{2\sqrt{\pi}}\left[ P_{2}+\frac{2P_{3}}{\sigma+1}(\frac{\rho}{\rho_{0}})^{\sigma-1}+P_{4}(\frac{\rho}{\rho_{0}})^{\gamma-1}\right]g(k) 
\end{equation}
from which the nucleon effective mass  for symmetric matter can be
evaluated according to
\begin{equation}
m^{*}_{r}\equiv\frac{m^{*}}{m_{0}}=\left[1+\frac{m_{0}}{\hbar^{2}k}\frac{\partial U}{\partial k}\right] ^{-1}_{k=k_{F},\rho_{0}} 
\end{equation}

\noindent
In the aforementioned framework of the MF approximation, 
all the quantities characterizing the reference  functional 
can be obtained in a relatively simple way by
solving a linear system with the strength parameters as unknown quantities.
Different branches of solutions can be obtained in correspondence  to
different values of $\sigma$ and $\gamma$ parameters. They define the shape of the different density form factors and  can also be relevant  in determining the slopes at density far from the saturation one.
For the three chosen examples the values of these quantities are
reported in Table 1.
\begin{table*}[htbp]
\caption{\label{tab:table1}%
For the three investigated cases we report the strength parameters values 
characterizing the adopted effective interaction, the chosen values of $\sigma$ and $\gamma$ and the resulting value of the slope of the symmetry energy at saturation L (see text).
The units of the $P$ and $L$ parameters are MeV. All the cases corresponds 
to $m^{\ast}_{r}$ and $m^{\ast}_{nr}-m^{\ast}_{pr}$ equal to 0.67 and $0.4\beta$}
\vskip 10pt
\begin{tabular}{cccccccccc}
\hline
\textrm{$P_{2}$}&
\textrm{$P_{3}$}&
\textrm{$P_{20}$}&
\textrm{$P_{30}$}&
\textrm{$P_{\pi}$}&
\textrm{$P_{40}$}&
\textrm{$P_{4}$}&
\textrm{L}&
\textrm{$\sigma$}&
\textrm{$\gamma$}\\
\hline
1042.8&-434.1&-870.0&169.9&-213.4&478.1&-300.0&63.3&0.9&0.7\\
368.3&55.9&-490.0&0.82&-198.9&289.4&-150.0&50.6&1.4&0.5\\
601.7&-19.1&-860.0&187.3&-212.7&477.8&-300.0&81.3&1.2&0,8\\      
\end{tabular}
\end{table*}

Finally, in the upper panels of Fig. 1 we show for the reference case associated
to the parameters reported in the first row of Table 1
the total energy as a function
of the relative density $\rho_{r}=\dfrac{\rho}{\rho_{0}}$ and in the bottom panels the related
symmetry energies. The results are plotted by means of a red line.
\subsection{Effective interaction in the molecular dynamics approach.}
Usually in quantum molecular dynamics approaches(see as examples Ref. \cite{tra2022,massnp,mdicozma})  the effects related to
the finite range interaction are  taken in to account by considering the momentum dependent interaction evaluated from the corresponding MF limit. This
can be also considered as a kind of approximate way to take into account exchange terms produced in an anti-symmetrized dynamics. 

In quantum molecular dynamics approaches single particles wave functions are represented through WPs with fixed width ($\sigma_{r}$=1.15 fm in CoMD \cite{comd,bon2}).
\begin{eqnarray}
\phi^{i}(\textbf{r})=\dfrac{1}{(2\pi \sigma_{r}^{2})^{3/4}}
\textit{e}^{-\dfrac{(\textbf{r}-\textbf{r}_{0,i})^{2}}{4\sigma_{r}^{2}}+i\textbf{k}_{0,i}\textbf{r}}
\end{eqnarray}
Therefore
the local terms  and the MDI ones 
are here obtained through a convolution operation 
with the WPs.

The 2-body MDI contribution associated to the generic couple $i$,$j$ is:

\begin{align}
E_{2,i,j}^{MDI}& \equiv \dfrac{1}{2}P_{2}R_{i,j}^{MDI}\\
R_{i,j}^{MDI}&=-\dfrac{1}{8\sigma_{r}^{3}}\xi^{3}
\\
&&\times\textit{exp}{-\dfrac{1}{4}\left[\dfrac{(\textbf{r}_{0,i}-\textbf{r}_{0,j})^{2}}{\sigma_{r}^{2}}+\xi^{2}
(\textbf{k}_{0,i}-\textbf{k}_{0,j})^{2}\right]} 
\nonumber
(\delta_{\tau_{i}-\tau_{j}}\delta_{s_{i}-s_{j}})
\end{align}  
with
$\dfrac{1}{\xi^{2}}=\dfrac{1}{4\sigma_{r}^{2}}+\dfrac{1}{\mu^{2}}$. $\tau_{i}$ and $s_{i}$ indicate 
the nucleon third components of the iso-spin and spin quantum numbers
respectively.
We note that the 
convolution of the finite range interaction with
the  WPs for the generic couple of identical particles affects the widths of the different overlap integrals. 
In fact the square of  the width in momentum space $\frac{1}{\xi^{2}}$ is just the quadratic sum of the widths associated to the MDI  interaction in MF approximation and  the width of the Wigner transform in momentum-space.
  
The  2-body direct term  associated to the finite range interactions will be:
\begin{align*}
E_{2,i,j}^{dr}\equiv \dfrac{P_{2}}{2}R_{2,i,j}^{dr};
\hskip 8pt
R_{2,i,j}^{dr}=\dfrac{1}{8\sigma_{r}^{3}}\xi^{3}\textit{e}^{-\dfrac{(\textbf{r}_{0,i}-\textbf{r}_{j})^{2}}{\lambda^{2}}}
\end{align*}
where $\lambda^{2}=4\sigma_{r}^{2}+\mu^{2}$ is still a quadratic width
obtained as composition of the interaction and WP widths in the configuration space.
By expressing, the total overlap
integral for the generic particle as:
\begin{eqnarray}
S_{v}^{i}=\sum_{j \neq i,j=1}^{A}\dfrac{1}{(4\pi \sigma_{r}^{2})^{3/2}}
\textit{e}^{-\dfrac{(\textbf{r}_{0,i}-\textbf{r}_{j})^{2}}{4\sigma_{r}^{2}}}
\end{eqnarray}
the finite range terms for the three-body and the iso-vectorial interactions are expressed as:
\begin{align}
E_{3,i,j}&=\dfrac{P_{3}}{\sigma+1}(\dfrac{S_{v}^{i}}{\rho_{0}})^{\sigma-1}[R_{i,j}^{MDI}+R_{i,j}^{dr}]\\
E_{4,i,j}&=\dfrac{1}{2}P_{4}
(\dfrac{ S_{v}^{i}}{\rho_{0}})^{\gamma-1}[R_{i,j}^{MDI}
+(2\delta_{\tau_{i}-\tau_{j}}-1)R_{i,j}^{dr}]
\end{align}

Finally the ZR contributions will give:
\begin{align}
E_{i,j}^{20}&=\dfrac{1}{2}P_{20}R_{i,j}^{0}\\
E_{i,j}^{30}&=\frac{P_{30}}{(\sigma+1)}\frac{(S_{v}^{i})^{(\sigma-1)}}
{\rho_{0}^{\sigma}}R_{i,j}^{0}\\
E_{i,j}^{\pi}&=\dfrac{P_{\pi}}{2}\frac{ (S_{v}^{i})^{(\gamma-1)}}{\rho_{0}^{\gamma}}R_{i,j}^{0}
\delta_{s_{i}+s_{j}}\\
E_{i,j}^{sy}&=\dfrac{1}{2}P_{40}
\dfrac{ (S_{v}^{i})^{(\gamma-1)}}{\rho_{0}^{\gamma}}
[(2\delta_{\tau_{i}-\tau_{j}}-1)R_{i,j}^{0}]\\
R_{i,j}^{0}&=\dfrac{1}{(4\pi \sigma_{r}^{2})^{3/2}}
\textit{e}^{-\dfrac{(\textbf{r}_{0,i}-\textbf{r}_{j})^{2}}{4\sigma_{r}^{2}}}
\end{align}

\noindent
By adding these contributions for all the couples of nucleons
and adding the kinetic terms related to the Fermi motion we obtain the total energy expressed through the centroids of the WPs.
As happens for all the QMD-like models with fixed widths WPs, the time evolution of the WP centroids will be obtained by solving the classical equation of motion associated to the obtained Hamiltonian. 
Quantum effects related to the Pauli principle are introduced through the blocking factors for nucleon-nucleon collisions and in CoMD model also through the usage of impulsive forces 
associated to the numerical  Pauli constraints \cite{comd,tra2022} (see also next
sections).
\noindent
Finally, we observe that from the previous expressions for the MDI interaction it is possible to obtain an approximate form
for the relative effective mass in MD approaches, $\widehat{m_{r}^{*}}$, 
formally obtained from Eq. (23) with the following substitutions/association:
$\mu\rightarrow \xi$, the generic overlap integral $S_{v}^{i}\rightarrow \rho$ and by dis-coupling the nucleon-nucleon phase-space coordinates (see Eq. (27)) in the necessary integration operation leading to the Eq. (23). 
But in the $\widehat{m_{r}^{*}}$ values the effects
of the correlated fluctuations produced by the dynamcal model are neglected.

\subsection{Source of many-body correlations in phase-space}
Before discussing the calculations performed with the CoMD model, it's worth highlighting the source of correlations contained in the molecular dynamical model strictly connected to the interaction.

\noindent
With reference to the MDI interaction,
it is possible to express  in general terms the total potential energy per nucleon of our system in the MF limit as follows:
\begin{equation}
E_{MF}=-\dfrac{P_{2}A}{2}\sum_{in}\overline{\overline{A^{MF,in}(\textbf{r},\textbf{r'})B^{MF,in}(\textbf{k},\textbf{k'})}}
\end{equation}
where A is the number of nucleons.
and $in$ indicates  the different kinds of interactions.
Hereinafter we will assume A as very large but finite. 
In the following this assumption (very large volume) allows formally to substitute with a very good approximation the  integration on  the particle momentum coordinates in the MF expression with a sum on the particle indexes. The discrete summation on wave-packet coordinates is instead more suitable for the MD models. 
The double overline symbol represents 
the average over the number of nucleon couples i.e.
\begin{equation}
\overline{\overline{A^{MF,in}(\textbf{r},\textbf{r'})B^{MF,in}(\textbf{r},\textbf{r'},\textbf{k},\textbf{k'})}}=
\sum_{i\neq j}\dfrac{A_{i,j}^{MF,in}(\textbf{r}, \textbf{r'})B_{i,j}^{MF,in}(\textbf{k},\textbf{k'})}{A^{2}}
\end{equation}

\noindent
In the following we want to make explicit the form of the functionals $A$ and $B$ in the MF limit, as an example we refer to one of the different terms i.e.
to the momentum dependent part of the two-body interaction (see Eqs. (8,9)). 
We assume that all the A nucleons are identical.
By performing the summations on the nucleons we get the average potential energy per nucleon:
\begin{equation}
E_{MF,2}=-\dfrac{P_{2}A}{2V}(\sqrt{\pi}\mu)^{3}\sum_{i\neq j}
\dfrac{\textit{e}^{-(\textbf{k}_{i}-\textbf{k}_{j} )^{2}\mu^{2}/4}}{A^{2}}
\end{equation}
Apart from the pre-factor $\dfrac{1}{2}$
(it avoids double counting in the total interaction), 
it can be noted that the generic term of the above sum corresponds to
the one already indicated in Eq.(9) in which the dependence on the spacial coordinates has been
absorbed through the convolution with plane waves.
With respect to the previous expression  the following
associations can be made:

\begin{equation}
\overline{\overline{A^{MF,2}}}=\dfrac{1}{V}=A^{MF,2}_{i,j}
\end{equation}
\begin{equation}
B^{MF,2}_{i,j}=(\sqrt{\pi}\mu)^{3}\textit{e}^{- (\textbf{k}_{i}-\textbf{k}_{j})^{2}\mu^{2}/4} 
\end{equation}

\begin{equation}
\overline{\overline{B^{MF,2}}}=\dfrac{1}{A^{^{2}}}(\sqrt{\pi}\mu)^{3}\sum_{i\neq j}
\textit{e}^{-(\textbf{k}_{i}-\textbf{k}_{j} )^{2}\mu^{2}/4}
\end{equation}

Through these associations in the MF case we obtain that: 
\begin{equation}
E_{MF,2}=-\dfrac{P_{2}}{2}A \overline{\overline{A^{MF,2}}}\centerdot \overline{\overline{B^{MF,2}}}
\end{equation}

\noindent  
For the MD case we can proceed in strict analogy by substituting plane-waves  
with the WPs in to the matrix Eq.(6). 
Then after algebraic transformations the following expression is obtained :

\begin{align*}
E_{MD,2}^{i,j}=-\dfrac{P_{2}}{2}\dfrac{1}{(2\pi\sigma_{r}^{2})^{3}}
\int_{V_{x}}\int_{V}\textit{e}^{i(\textbf{k}_{i}-\textbf{k}_{j})\textbf{x}}
\textit{e}^{-\textbf{x}^{2}/\xi^{2}}\textit{e}^{-(\textbf{r}_{i}-\textbf{r}_{j})^{2}/4\sigma_{r}^{2}}\\
\textit{e}^{-\textbf{R}^{2}/4\sigma_{r}^{2}}d^{3}xd^{3}R
\end{align*}
after the volume integrations we get:
\begin{equation}
E_{MD,2}^{i,j}=-\dfrac{P_{2}}{16}\dfrac{\xi^{3}}{\sigma_{r}^{2}}
\textit{e}^{ -(\textbf{r}_{i}-\textbf{r}_{j})^{2}/4\sigma_{r}^{2}}
\textit{e}^{-\xi^{2}(\textbf{k}_{i}-\textbf{k}_{j})^{2}/4}
\end{equation}
(see also Eqs. (26-27)).
In this case the spacial localization of the WPs do not allow for the averaging effect above discussed and obviously the phase-space wave packed centroids appearing in Eq. (43) will produce strong dynamical  correlations through the equations of motion.
 For the potential energy per nucleon we obtain:
\begin{equation}
E_{MD,2}=-\dfrac{P_{2}}{2}A\sum_{i\ne j}
\dfrac{1}{(4\pi\sigma_{r}^{2})^{3/2}A^{2}}
\textit{e}^{ -(\textbf{r}_{i}-\textbf{r}_{j})^{2}/4\sigma_{r}^{2}}
\pi^{3/2}\xi^{3}\textit{e}^{-\xi^{2}(\textbf{k}_{i}-\textbf{k}_{j})^{2}/4}
\end{equation}
Similarly to the MF case we can perform the following associations:

\begin{equation}
A^{MD,2}_{i,j}=\dfrac{1}{(4\pi\sigma_{r}^{2})^{3/2}}
\textit{e}^{ -(\textbf{r}_{i}-\textbf{r}_{j})^{2}/4\sigma_{r}^{2}}
\end{equation}

\begin{align}
\overline{\overline{A^{MD,2}}}=\sum_{i \neq j}\dfrac{1}{(4\pi\sigma_{r}^{2})^{3/2}A^{2}}
\textit{e}^{ -(\textbf{r}_{i}-\textbf{r}_{j})^{2}/4\sigma_{r}^{2}}
\end{align}

\begin{align}
\overline{\overline{B^{MD,2}}}=\sum_{i \neq j}\dfrac{1}{A^{2}}\pi^{3/2}\xi^{3}\textit{e}^{-\xi^{2}(\textbf{k}_{i}-\textbf{k}_{j})^{2}/4}
\end{align}
\begin{align}
B^{MD,2}_{i,j}=\pi^{3/2}\xi^{3}\textit{e}^{-\xi^{2}(\textbf{k}_{i}-\textbf{k}_{j})^{2}/4}
\end{align}
From the previous expressions we note that in the MD case the factorization expressed in Eq. (42) is not obtained so that:
\begin{align}
E_{MD,2}=-\dfrac{P_{2}}{2}A \overline{\overline{A^{MD,2}B^{MD,2}}}
\end{align}
where for simplicity of notation we have eliminated  the coordinates $\textbf{r},\textbf{r'}$,$\textbf{k},\textbf{k'}$ as arguments of the functional.
It is possible now to highlight the differences 
with respect to the MF case. To this end we can define the following quantities that represent
a kind of dynamical "fluctuations" with respect to the MF quantities:
\begin{equation}
\delta A_{i,j}^{MD,2}=A_{i,j}^{MD,2}-
\overline{\overline{A^{MF,2}}};\hskip 10pt
\delta B_{i,j}^{MD,2}=B_{i,j}^{MD,2}-
\overline{\overline{B^{MF,2}}};
\end{equation}
Using the above expressions we can now rewrite the MD functional as:
\begin{equation}
E_{MD,2}=E_{MF,2}+\delta E_{b}
\end{equation}

\begin{equation}
\delta E_{b} =-A\dfrac{P_{2}}{2}(\overline{\overline{\delta A_{i,j}}}
\centerdot\overline{\overline{B^{MF,2}}}+\overline{\overline{\delta B_{i,j}}}\centerdot\overline{\overline{A^{MF,2}}}+\overline{ \overline{\delta A_{i,j}\delta B_{i,j}}})
\end{equation}
where $\delta E_{b}$ indicates the correction on the
binding energy per nucleon. 
\noindent
Beyond the effects due to the many-body dynamics the first two contributions in Eq. (52) can be largely  affected
by the convolution operation with the Gaussian associated to the WPs (differences in the widths of the overlap integrals with respect to the MF case).
The last term
will generate the  dynamical correlation between the "fluctuations" in phase-space associated to the functional $A^{in}_{i,j}(\textbf{r},\textbf{r'})$ and  $B^{in}_{i,j}(\textbf{k},\textbf{k'})$ with respect to the mean value related to the MF approaches (see also
Sec. 2.5).
\noindent
More generally, we now have to observe that   another source of correlations able to produce
a $\delta E_{b}$ different from zero is obtained
whenever a microscopic interaction (also for zero range) acts in a different way, for example with opposite signs, on two different subsets of nucleon couples.
In this case  the interaction can affect
on average the overlaps integrals related to the two subsets (i.e. can affect the relative distances between  the nucleons belonging to the different kinds of couples) modifying the related energies. 
This pure two-body effect cannot be achieved in the MF case just due to the one-body internal average generated by the use of plane-wave.
An example of these correlations  was studied in
\cite{miosky}.  This was the case of the iso-vectorial interaction coupled with the Pauli constraint affecting in different way $nn$ and $pp$ couples with respect to the $np$ ones. This kind of short range correlations
can be interpreted as a tendency to form deuteron-like particles (or more generally clusters) also at density 
around the saturation one (see also Appendix B for cluster production at low density).
This effect will be highlighted in the next sections also in the
case of the momentum dependent interaction.
\noindent

 \subsection{Box calculations with CoMD model}
 Before to discuss the results obtained from box calculations we shortly  comment about the choice of $\sigma_{r}=1.15$ fm parameter in the CoMD model. 
 As in QMD-like model $\sigma_{r}$ does not change in time and it fixes the width of the WPs. 
 According to Ref. \cite{comd} and to the ZR Skyrme interaction adopted in the same work, this parameter
 was chosen  in such away to reproduce within a precision of about 10\% good "ground state" properties (root-mean square radius, binding energies)
 for  nuclei with mass in the range 40-210 .
 $\sigma_{r}$ determines also the magnitude of the fluctuations and in the CoMD model it determines also the scale of the coalescence radius $R_{c}\simeq 2\sigma_{r}$ used to identify the formation of clusters. 

\noindent
Now it is worth noting that in the limit of $\sigma_{r}\gg\mu$ that is: $\xi\rightarrow\mu$ and $S_{v}^{i}\rightarrow \rho$ the averaging effect of one body mean-field is completely restored and therefore the expressions in Sect. 2.1 have to coincide with the ones displayed in Sect. 2.2.
 This limit has been  used as a numerical test for the CoMD algorithm associated to the new effective interaction.

\noindent
After this check,  the expressions Eqs. (26-35) which are functions of the wave-packed centroids have been evaluated in box calculations
with periodic boundary conditions to simulate the "ground state" NM (see  Appendix A). 
In this stage the set of  MF strength parameters reported on the first row of Table 1 was used.
The calculations have been performed for different 
densities (number of particles in the unitary volume) 
and $\beta$ values.
In  Appendix B we show  analogues calculations using a ZR interaction.
 
\noindent 
2000 WPs  with centroids distributed in a box of different sizes were considered according to different densities. For each different
 neutron-proton combination and density, four different microscopic realizations have been obtained by uniformly distributing neutrons and 
protons in the cubic box having momenta  inside the related Fermi spheres.  Half number of the  neutrons and protons will have spin-up and the other half spin-down. 
\noindent
The time evolution of the WP's was applied for each configuration following the CoMD approach \cite{comd}.
The results obtained  for the total energy is the average on the four independent configurations. 
\noindent
For sub-saturation densities
the  WPs coordinates were not evolved in time.
Only the pre-stabilization (PrS) stage is applied  (see Appendix A) in which only the Pauli constraint act.
This is because  the increasing of fluctuations 
and the clusterization processes
strongly affect the homogeneity-uniformity conditions 
at low density, which instead are reflected in the static MF expression given by Eqs. (5-15). 
To avoid discontinuity in the obtained functional,
the time evolution is applied  starting from the saturation values (a relatively large value for the present case). 
In fact at $0.9-0.8\rho_{0}$ the evaluation of the total energy using only the PrS stage or the PrS and the time evolution gives within the 3-6\% the same result.
The low density interesting case \cite{bon1,pasta}   certainly deserves a separate study.  In  Appendix C it is  only briefly  illustrated how the constraint can affect the primary light cluster production at a density equal to $0.2\rho_{0}$.

\noindent
In the upper panel of Fig. 1, for $\beta=0$, are shown with open symbols the obtained results from CoMD calculations concerning the total energy as a function of the relative density $\rho_{r}$ for the chosen  case.
The error for each determination is within the symbol size.
The red solid lines represent instead the values  obtained 
from  the MF prediction which accurately reproduce the chosen characterizing EDF reference properties reported at the beginning of  Sect. 2.

\noindent
Large values of
$\delta E_{b}$ derived from the comparison can be pointed out.  
The amount of $\delta E_{b}$  clearly depends also
on the chosen values of some parameters such as the nucleon effective masses for symmetric and asymmetric NM , the stiffness of the iso-vectorial contribution etc.. 
For instance,  effective mass values (in the MF limit) more close to 1   and $\gamma \simeq 0.5$ for the  iso-vectorial interaction decrease the value of $\delta E_{b}$ to about 2.5 MeV.
But for the typical set
of saturation EoS properties used in these examples (Table 1)  the
energy shift can reach value as large as 6 MeV.

\noindent
In the bottom panel, for the same case, open symbols show the values obtained for the symmetry energy.  Positive differences in regard to the  MF case are relatively small for density around  the saturation value (the increasing is of the order of 5-10\%). They  slightly increase at lower density.
This enhancement in regard to the MF reference is due to a slight increase of the overlap in $np$ couples with respect to the other couples due to the iso-vectorial
interaction and to the Pauli constraint \cite{miosky}.   This mechanism is also discussed in  Appendix B.

\begin{figure}[htbp]
\includegraphics[width=5.in]{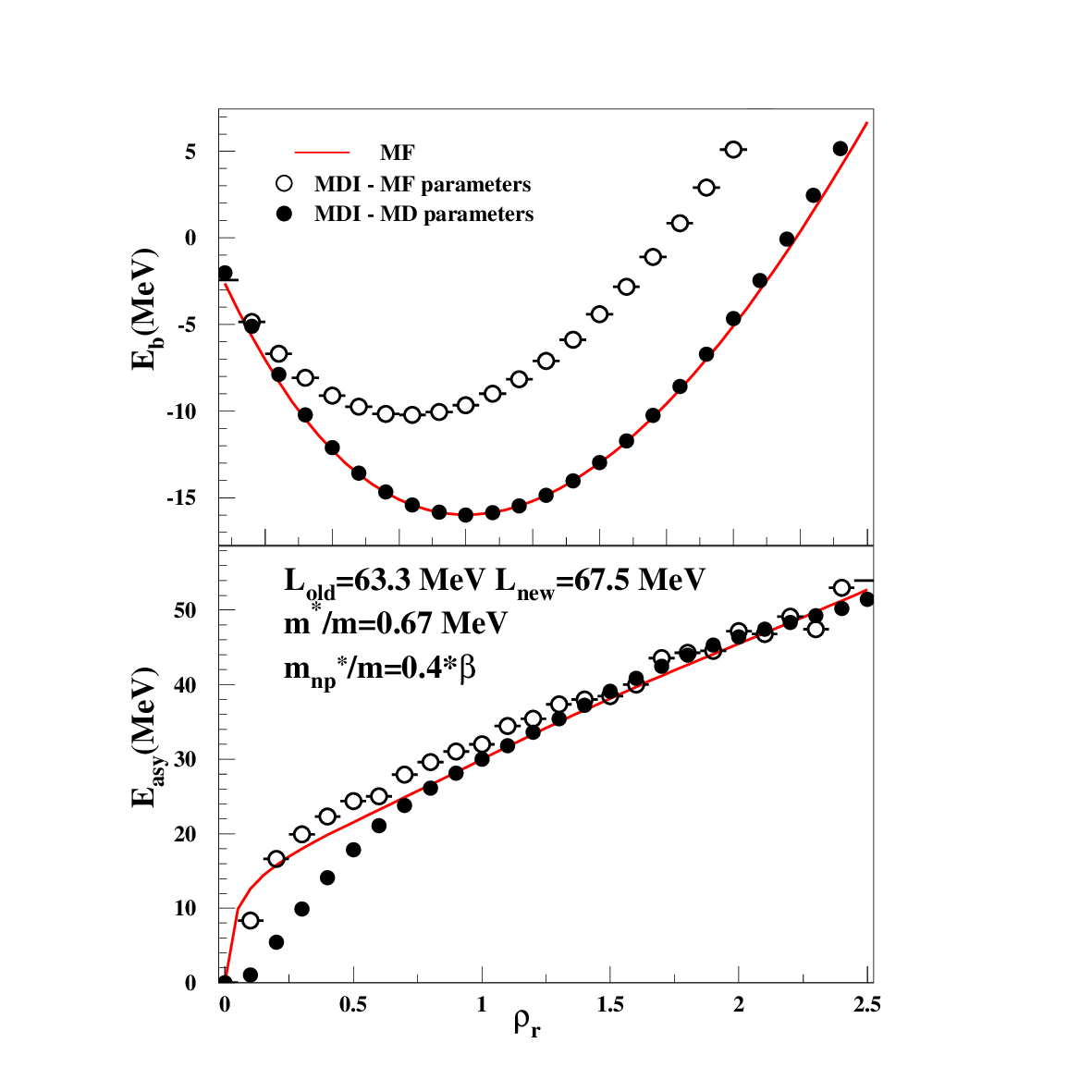}
\caption{\label{Fig:1} For symmetric NM, in the upper panel  with a red line the total energy per nucleon $E_{b}$ is shown as a function of the relative density $\rho_r$.  
It is evaluated in the MF approximation using the parameters shown in the first row of Table 1. 
In the bottom panel the corresponding values of the symmetry energy are also shown.
 For the upper panel and lower panel
 the open symbols corresponds to results obtained through CoMD box calculations performed  using the same set of values for the strength parameters in the first row of Table 1 (MDI-MF).
The black dots are the CoMD results obtained with the new set of values shown in the first row of Table 2 (MDI-MD). 
 Errors are inside the size of symbols.
 (color)} 
\end{figure}

\subsection{Discussion of the obtained results}
In this section an analysis from a qualitative and 
quantitative point of view the of aforementioned differences is illustrated.

\noindent
For the case already reported in Fig. 1, in Fig. 2 (upper panel) we plot at the 
saturation density $\rho_{0}$, the average Pauli over-blocking
$F_{P}=\overline{f}-f_{m}$ as function of the number of steps  
$N_{s}$ that define the variable $t_{s}=0.25*N_{s}$.
The initial configuration is obtained by distributing uniformly and randomly
the WP centroids in the available phase-space given by 
the Fermi sphere and the volume of the main cell.
$\overline{f}$ represents the average occupation in the phase-space in 
a volume $h^{3}$
This volume has been choosen as representative of WPs occupation in phase-space.  
while $f_{m}$
is a threshold minimum value associated with the numerical uncertainty 
of the constraint method used in the present CoMD calculations (see Appendix A).
In this stage (PrS) the coordinates are not evolved in time but only the
numerical procedure related to the constraint is applied
(see Appendix A for details). 

\noindent
In the bottom panel  the corresponding value of the total energy is plotted. We see that already for $t_{s}=0$ the energy is larger
than the expected value of -16 MeV as already shown in Fig. 1. This is due to the convolution with
the WPs in the evaluation of the effective interaction.
Moreover as a function of $t_{s}$, the energy still increases up to a level  of about 1.5 MeV in a correlated
way with the reduction of $F_{P}$.
\begin{figure}[htbp]
\includegraphics[width=5.in]{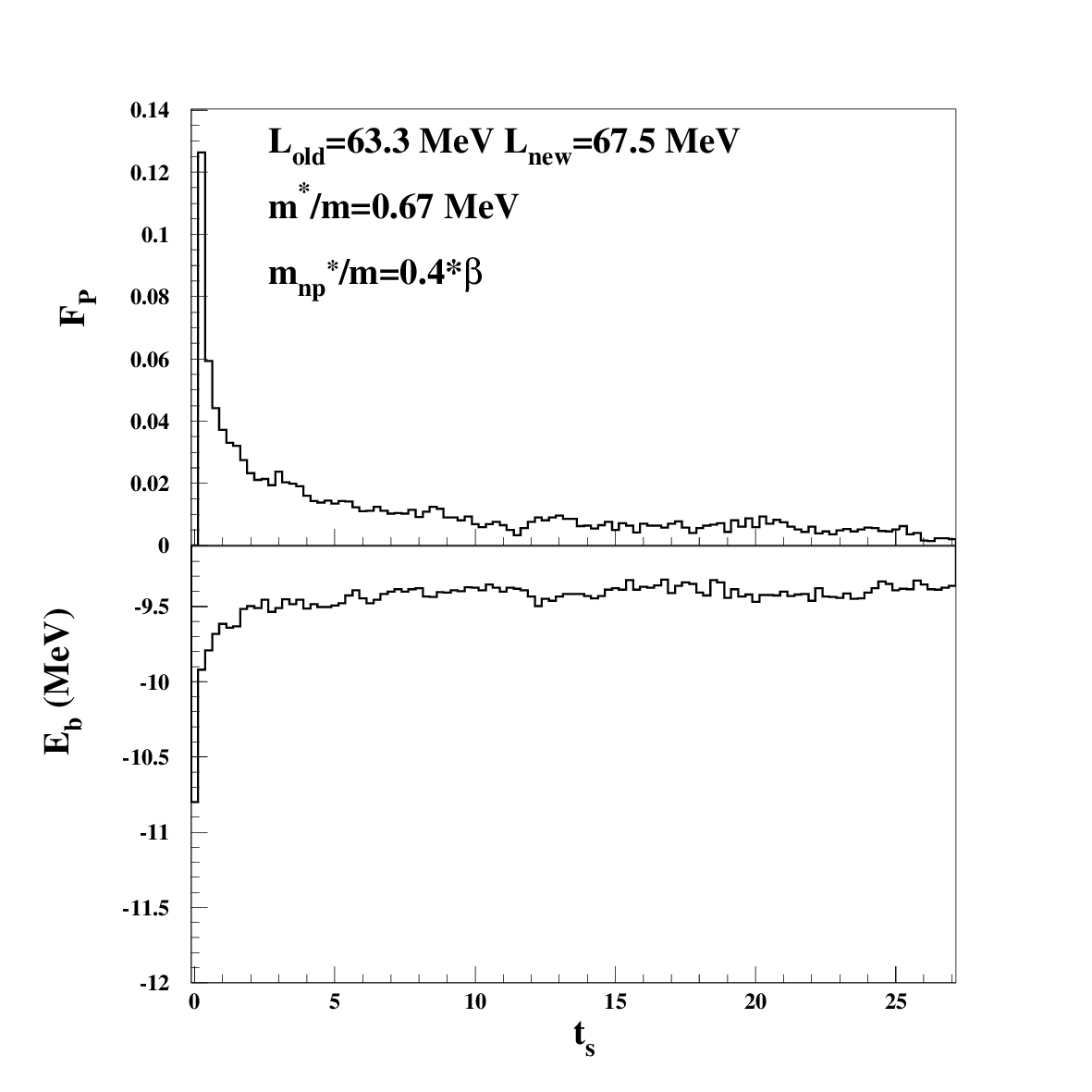}
\caption{\label{Fig:2} In the upper panel the Pauli-over-blocking $F_{P}$ is plotted
as a function of the  time step $t_{s}$(see text and Appendix A). The box calculations are performed 
at the saturation density  with an effective interaction corresponding to the parameters 
reported in the first row of TABLE I.
In the bottom panel the corresponding total energy $E_{b}$ per nucleon is plotted.}
\end{figure}

\noindent
This further increment of the total energy  may be considered as a configuration energy associated to the correlation in  phase space
induced by  the Pauli constraint. In fact, in the analyzed case,
the total  potential energy arising from the finite range interaction  is negative,  and the CoMD constraint tends to minimize the number of couples of identical nucleons which are near in phase space where 
this  negative MDI term is  more effective.  
The PrS stage leading to a minimization of the average Pauli blocking
represents a typical stage in CoMD calculations applied to 
finite systems and it's used in the procedure aimed at searching for the more stable "ground state" configurations. For the further evolution the constraint is applied at each time step of the collision processes  hindering
the spontaneous evolution of the phase-space distribution to a classical Boltzmann one \cite{comd}.
In  Appendix A are described both the PrS in some detail  and the time evolution of box calculations after the PrS including the procedure adopted to obtain a kind of "ground state" for the NM simulation.

\noindent
Therefore summarizing the result shown in Fig. 1, it can be said that
for values of the nucleon effective mass significantly different from 1, 
the energy correction $\delta E_{b}$ related to the usage of WPs  and  to the system Fermionic dynamics can be rather  large (see also Sec. 2.3). 

\noindent
In the following the procedure to obtain the chosen reference EoS is shortly described.
In  Appendix B for the sake of completeness, analogue calculations to the one presented in this section are illustrated for the case
of ZR interaction.  

\subsection{The search for the new set of parameter values}
The calculations illustrated in the previous section have
been performed for different densities and charge-mass asymmetries.
The average overlaps $R^{MDI}$ and $R^{dr}$ and the ones related to the local interaction   
have been fitted with 
sixth order 
polynomials of the density
and for each density with fourth order polynomials of the asymmetry parameter $\beta$.
It's worth noting that the average overlap integrals, as evaluated from the 
described procedure in the previous section contain  all the main correlations effects associated to 
the CoMD phase-space constraints related to the "ground state"  configurations.

\noindent
The quality of the best-fit was rather satisfactory in a wide range of densities around the saturation one. 
With this new functional of the density,
it has been possible to define a linear system in the new parameters 
 $P'_{n}$  by imposing the usual conditions (except for  the ones related to the effective masses) already expressed in Sec.II concerning symmetric and asymmetric NM ground state properties.
Table 2 contains the new set of obtained parameter values for the cases evaluated in this work. Their uncertainty is of the order on $\pm$3\%. 
The value of the total energy  is plotted in Fig. 1 with
black circles for the $\gamma=0.7$ case.
\begin{table*}[htbp]
\caption{\label{tab:table2}%
In the table we report the set of new parameters  $P'$ for the different values of $\gamma$ and $\sigma$ as obtained from the fit procedure.
The units of the $P'$  parameters are MeV.
The slope parameter  $L$ of the symmetry energy from the first to the third row are
67.5 MeV, 55 MeV and 105.0 MeV respectively
The global uncertainty on the parameters values
as due to the model calculations and fit procedure is of the order of $\pm$3\%.}.
\vskip 10pt
\begin{tabular}{ccccccccc}
\hline
\textrm{$P'_{2}$}&
\textrm{$P'_{3}$}&
\textrm{$P'_{20}$}&
\textrm{$P'_{30}$}&
\textrm{$P'_{\pi}$}&
\textrm{$P'_{40}$}&
\textrm{$P'_{4}$}&
\textrm{$\gamma$}&
\textrm{$\sigma$}\\
\hline
1776.8&-1279.6&-1503.1&920.8&-171.4&551.1&-425.6&0.7&0.9\\      
450.8&-182.2&-639.0&190.3&-175.7&313.5&-213.1&0.5&1.4\\
660.8&-225.4&-1069.7&602.3&-220.3&684.8&-420.6&0.8&1.2\\
\end{tabular}
\end{table*}

\noindent
Finally the values of the approximated relative nucleon effective mass
$\widehat{m_{r}^{*}}$ introduced in Sect. 2.3 for symmetric matter are 0.82, 0.26, 0.84 for  parameter values in  Table 2 from row 1 to 3 respectively. 
A better estimation could be obtained  in a dedicate study starting from the microscopic configurations obtained in box calculations.

\noindent
The agreement with the reference case
(red line) is rather satisfactory.
In the bottom panel, the obtained symmetry energy fits in a satisfactory way the reference line around the saturation density and beyond.
At lower density the agreement is instead quite unsatisfactory. To improve the agreement, further conditions
associated with behaviour at the low density should be imposed
by letting to vary the $\gamma$ and the $\sigma$ parameters with respect the values of the reference case.

\section{Finite systems}
\subsection{ Finite Range Interaction
 and reaction Mechanisms}

In this section  the above discussed effective interaction in CoMD calculations is used to study, as an example, the $^{64}Ni+^{48}Ca$ collision at different incident energies. This system  has been also widely experimentally investigated  (see  Refs \cite{exp1,exp2,exp3,exp4} for example).   
It  has a relevant charge/mass asymmetry 
so that also the
iso-vectorial interaction plays a relevant role in the dynamics.
In particular, along with the almost charge/mass  symmetric partners like $^{58}Ni+^{40}Ca$, $^{64}Ni+^{40}Ca$  it allows  to experimentally investigate, with a good sensitivity,  the isospin equilibration phenomena 
through the disassembly of hot and dense intermediate systems(see as an example Refs. \cite{dip,exp4}).

A comparison between reaction mechanisms produced using a ZR effective interaction ($m^{*}_{r}=1$) with parameter values modified for the CoMD  correlations (ZR-MD)  (see also Appendix B) and the 
above-described finite range interaction (MDI-MD) will be illustrated in the next section.
In the same section we also add  the interesting comparison with the case of the MDI with parameters obtained in the MF scheme (MDI-MF). This comparison in fact allows to estimate the effects of
the corrections due to the model correlations on the dynamics of the heavy collisions.
Appendix B, for completeness, contains a comparison between results obtained for the ZR interaction with and without CoMD corrections, i.e.: ZR-MD and ZR-MF respectively.

\subsection{The interaction and the "ground state" configuration}
Moving on to finite systems the surface properties acquire a relevant role that cannot be described in the limit of an infinite NM.
Therefore a correction term $E_{s}$ is usually introduced in the 
total energy of A nucleons through the following expression
$E_{s}=\frac{C_{s}}{2}(4\pi \sigma_{r}^{2})^{1.5}\sum_{i=1,A}\nabla^{2}_{i}S^{i}_{v}$.
It's worth noting that this term, evaluated in box calculations with periodic boundary conditions,  takes negligible values in the calculations of the present work (less than $\pm 150$ KeV in the range $\rho_{r}=0.1-3$).   
In CoMD for finite systems, this correction term is fixed, trough a warming-cooling procedure  applied to the microscopic configuration of the system under study and driven through the Pauli constraint. It follows a stabilization stage.
The process is stopped when the system reaches the given binding energy, the root-mean-radius, average kinetic energy, in a stable way (within 8\% of the requested values) for a  time interval of some hundreds of fm/c. 

\noindent
In the following for the MDI-MD case are used the parameter values that fill the first row of Table 2. 
For the  ZR-MD case the strength parameters are instead listed in Appendix B. 
Finally for the MDI-MF case the parameters are written in the first row of Table 1. 

For  the three mentioned interactions good "ground state" properties are reached by setting  $Cs=-11$ MeV*fm$^{2}$,$Cs=3$ MeV*fm$^{2}$ and $Cs=-7$ MeV*fm$^{2}$ respectively.

\subsection{Selection and comparison of the reaction mechanisms and transverse flow}
Collisions in a wide range of impact parameters up to the "grazing" collision have been calculated with the CoMD model  
to identify different reaction mechanisms.
The nucleon-nucleon collision cross-sections and the algorithm to decide for the attempted collisions has been 
described according to Ref. \cite{bertrep}.
In these calculations  the constraint procedure 
does not include the charge-exchange algorithm which was activated in the PrS stage(see Appendix A)  of box calculations.
\noindent
In Fig. 3  for the case ZR-MD (see Appendix B) the bi-dimensional plot $Z-V_{p}$
for the collision $^{64}Ni+^{48}Ca$ at different incident energy is shown.
Z represents the charge of the fragment formed after a maximum time of 350 fm/c.
$V_{p}$ represents the velocity of the produced fragments in the laboratory frame along the beam direction. 
Fragments  are identified with a minimum-spanning-tree method.
The figures refers to an impact parameter range
$b=0-10.5$ fm.

\noindent
In order to testing the interactions, the present work will focus on the dynamics of processes producing 
fusion-incomplete fusion residues, fission fragments and intermediate mass fragments in the mid-rapidity region in central-semi-central collisions.
The aforementioned process are well localized in the
range of impact parameters $0-5$ fm.
For these central-semi-central collisions   these fragments in the mid-rapidity region are mainly produced through the formation of one intermediate system in which the largest changes of density are produced.
Furthermore, the primary source is formed by target and projectile   nucleons having on average a  relative velocity  directly established through the beam energy (participant zone). Therefore we think that  the aforementioned processes are better suited to discuss the behaviours of the interaction  as a function of the incident energy. 
Hence at this stage, the analysis of the more complex and diversified mechanisms induced by semi-peripheral collisions is not included.

\noindent
In particular in Fig. 3 the upper rectangles define the region where the fusion/incomplete fusion residues have been integrated.
Inside the lower rectangles will be evaluated the cross-section for production of Intermediate Mass Fragments (IMF)which are defined as fragments  having charge Z=3$\div$12. 
In the same region of velocity a kind of  "fission" of the hot residues is associated to the production, in the same event, of two fragments 1 and 2   heavier than the Nitrogen and having comparable charges (with the associate ratio of the charges Z1/Z2 greater than 0.8 and Z2 larger than or at most equal to Z1).
For the energies not shown in the figure the velocity ranges have been scaled proportionally to the center of mass (c.m.) velocity.
As observed before, in the present work the quantitative analysis will be focused on the impact parameter range $b=0-5$ fm where the above mechanisms are well localized.
\begin{figure}[htbp]
\includegraphics[width=5.in]{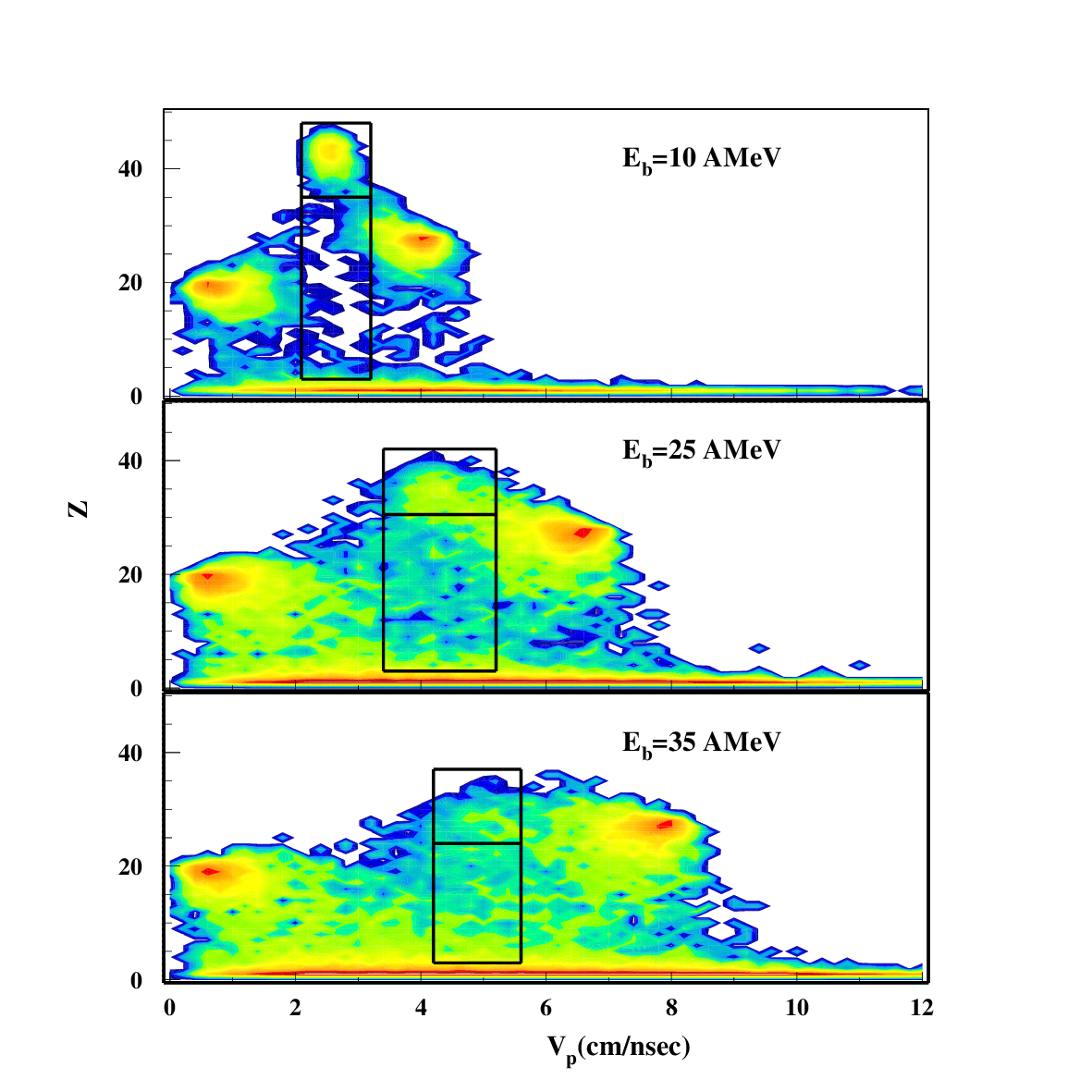}
\caption{\label{Fig:3} For the collisions 
$^{64}Ni+^{48}Ca$ (ZR-MD case), for the impact parameter range $b=0-10.5$ fm,  the
produced fragment charges Z versus the laboratory velocities $V_{p}$ along the beam direction are plotted for different incident energies.
The rectangles define the regions where the different reaction mechanisms are integrated (see text) (color).}
\end{figure}

\noindent
In the interval of considered impact parameters and for the energy interval 10-50 AMeV the mechanism is characterized by  the production of a bump in the $Z-V_{p}$  plots centred around the c.m. velocity.
In addition to the IMF production, heavy residues can be identified having charge Z higher than 20.
 This happens for both the MDI and the ZR cases cases.
As an example, the two upper panels (left and right) of Fig. 4 show such bumps for $E_{inc}$ equal to 10 and 35 AMeV. 
The last two bottom panels show the same plots 
at 100 AMeV and the bumps associated to the production of
an heavy residues are disappeared(see below).
Due to the impact parameter selection we note the absence of the binary mechanisms.
In Fig. 5 for the MDI-MD and ZR-MD cases, the integrated cross-section for the production of heavy residues (as  defined above) are shown as function of the incident energy.
They are represented by  red square and blue  
dots respectively.
In the inset panel the associated fission probabilities
estimated as: $P_{f}=\dfrac{N_{fis}}{N_{fus}+N_{fis}}$ are also plotted. In the above expression 
$N_{fis}$ and $N_{fus}$ indicate the selected number of events for fusion and fission respectively.

\begin{figure}[htbp]
\includegraphics[width=5.in]{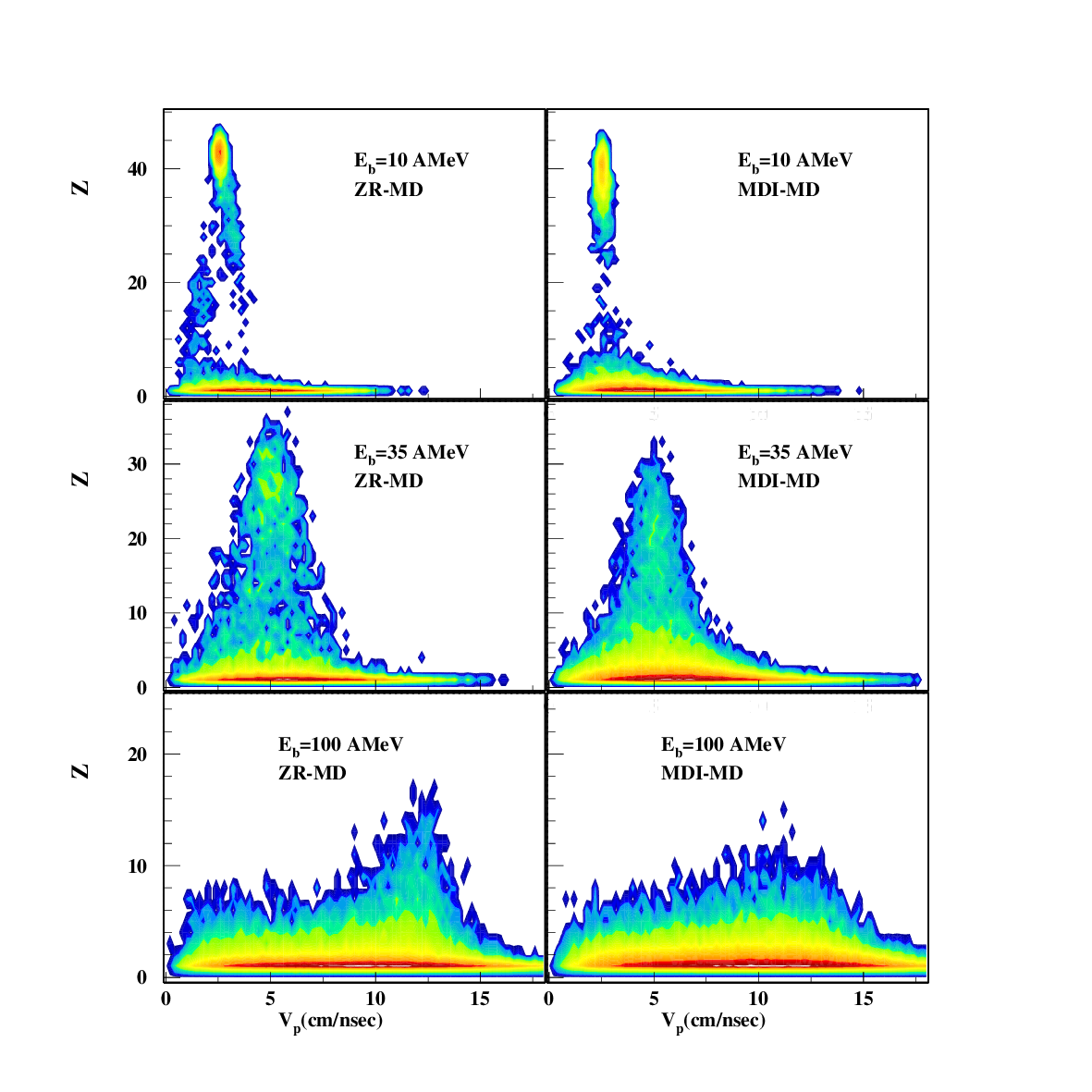}
\caption{\label{Fig:4}  $Z-V_{p}$ plots for the 
$^{64}Ni+^{48}Ca$ system. The different panels show the plot associated to the interactions ZR-MD and MDI-MD and to different energies as indicated in the figure. The range of impact parameters is $b=0-5$ fm.(color)}
\end{figure}
\noindent

\begin{figure}[htbp]
\includegraphics[width=5.in]{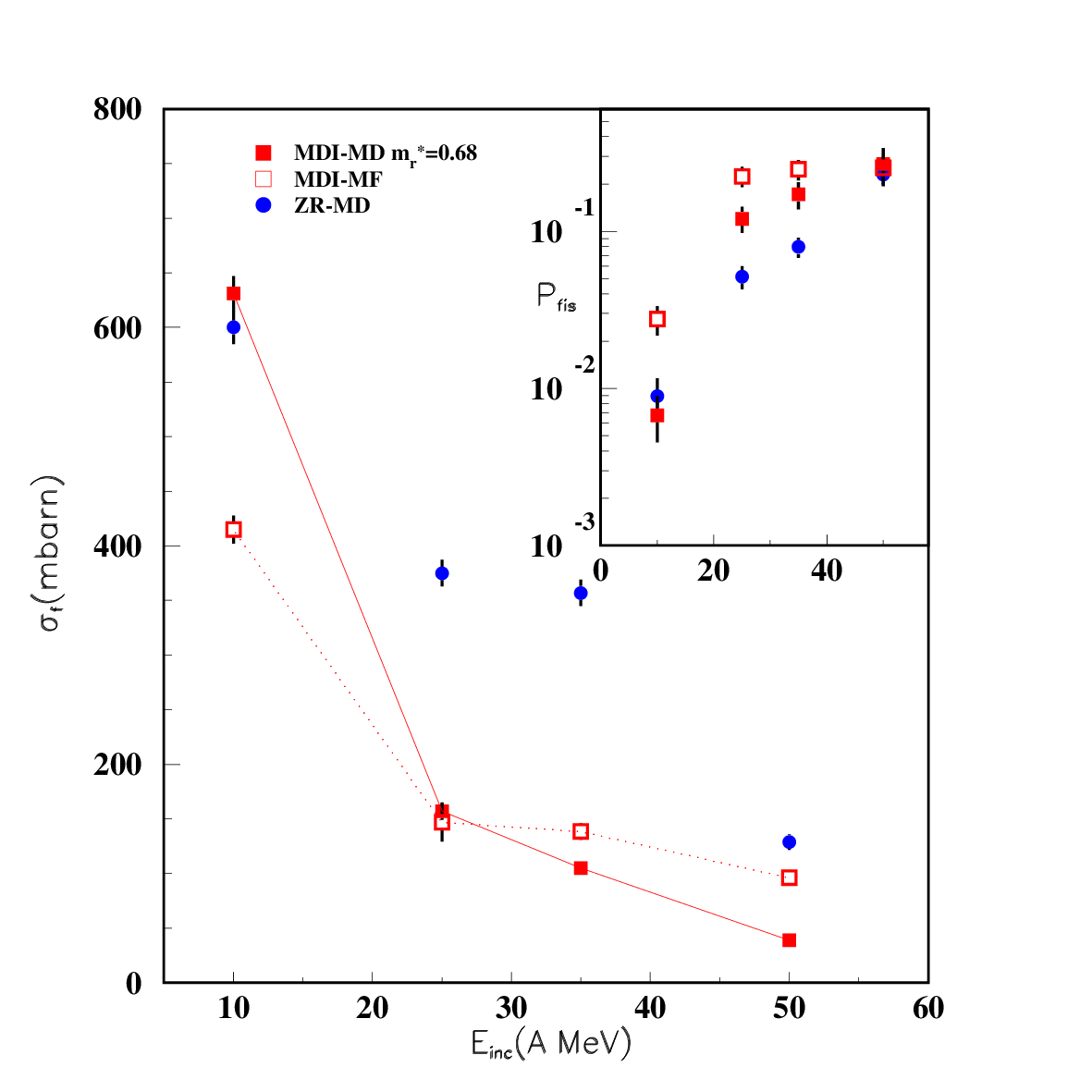}
\caption{\label{Fig:5}  As a function of the incident energy $E_{inc}$ for the $^{64}Ni+^{48}Ca$ collision and
for an impact parameter range $b=0-5$ fm  the evaluated cross-section associated the formation of an heavy residue is plotted.
The blue dots and the red square  symbols refer to the ZR-MD and MDI-MD cases respectively.
The open squares  represent the results for the MDI-MF case. 
In the inset the fission probabilities are also shown.
The lines joining the points MDI-MD and MDI-MF are plotted to simplify the comparison between the two cases. (color on-line)}
\end{figure}
\noindent

\begin{figure}[htbp]
\includegraphics[width=5.in]{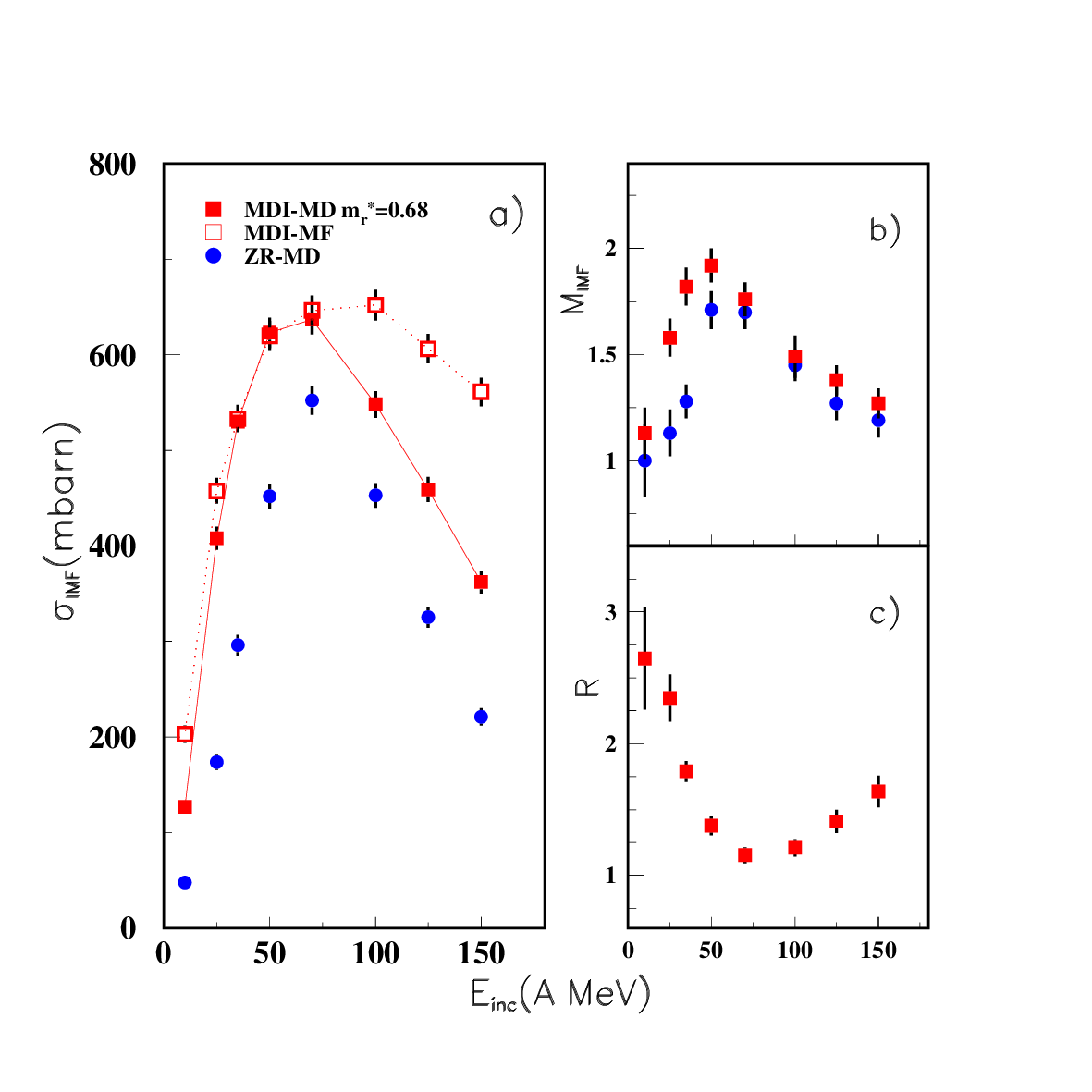}
\caption{\label{Fig:6}  For the system $^{64}Ni+^{48}Ca$, in the impact parameters range  $b=0-5$ fm: panel a) shows for the MDI-MD, MDI-MF, ZR-MD  cases the 
cross section for the production of at least one IMF  as a function of the incident energy $E_{inc}$, panel b) shows the
associated multiplicity MDI-MD and ZR-MD , panel c) shows the ratio
$R=\dfrac{\sigma_{IMF}^{MDI-MD}}{\sigma_{f}^{ZR-MD}}$.
The lines joining the points MDI-MD and MDI-MF are plotted to simplify the comparison between the two cases. (color on-line)}.
\end{figure}

\noindent
Within this energy range the globally more repulsive character of the finite range interaction (MDI-MD)case
gives rise to a more probable disassembly of the hot compound. It in fact produces a lower cross section for the 
formation of hot residues and an enhanced "fission" probability for  $E_{inc}$ at 25 and 35 AMeV. 
This more repulsive character of the MDI-MD is also confirmed through  the production of an higher rate of IMF. 
This is shown in Fig. 6 where the cross sections for producing at least one IMF in the mid-rapidity region and the related multiplicity are plotted in  panels a) and b).
We note that in both cases the rates of IMF have a maximum around 70 AMeV. The further
increase of the energy produces disassembly processes with an increasing
fraction of light particles with charge Z less than 3. 
This produces a lowering of the IMF rate in both  cases. 

\noindent
In both figures the open square symbols represent instead
the cross sections for the case MDI-MF.
The lines joining the points are plotted to simplify the comparison with the MDI-MD case. 
From this comparison, it can be clearly observed how the corrections on the values of strength parameters
due to the discussed MD correlations can sensitively affect the studied quantities.
The ZR-MD and the ZR-MF cases are compared in Appendix B.
Going back to the earlier comparisons MDI-MD, ZR-MD,
in the panel c) it is shown the ratio 
$R=\dfrac{\sigma_{IMF}^{MDI-MD} }{\sigma_{IMF}^{ZR-MD}}$ associated to the IMF production. 
The ratio $R$ in this first part decreases  with the energy.
For a further increase of the energy $R$ starts to increase. This behaviour can be explained by looking at the two bottom panels of Fig. 4.
For the ZR case  is clearly seen that already at 100 AMeV a bump at velocity  much higher than the c.m. one
appears (corresponding to $b\simeq 5$ fm). 
This could be associated with the onset  of a precursor mechanism for the production of project-like fragments. This bump is
also present for the MDI-MD fm case but it is less pronounced.
Therefore, the presence of this other mechanism produces around the c.m. velocity a larger depletion in the $Z-V_{p}$ plot for the ZR-MD case with respect to the MDI-MD one and justifies the increasing behaviour of the ratio $R$ shown in Fig. 4c  from 100 up to 150 AMeV.

\noindent
The  global repulsive action associated to the 
finite range interaction strongly affects also the nucleon transverse flow as shown in Fig. 7. As an example for $b=3$ fm, it produces the lowering of the balance energy (which it is instead seen for the ZR-MD case around 70AMeV)  with positive slopes in  the explored energy range. 
In the upper energy limits the slopes for the two cases become comparable
being more affected by the nucleon-nucleon collision rate.
\noindent
Finally we observe that both the fast decreasing trend of the ratio R in the first 70 AMeV  (corresponding to about 35 AMeV of relative motion) and the behaviour of the transverse flow slopes as a functions of the beam energy could be interpreted trough the existence of a characteristic relative energy per nucleon  $E \simeq\dfrac{\hslash^{2}}{2m_{0}}\dfrac{2}{\xi^{2}}$  (for symmetric system) beyond which the momentum dependent effects are attenuated. Here $\hbar\dfrac{\sqrt{2}}{\xi}$ establish the characteristic width of the MDI (see Eq.27).

\begin{figure}[htbp]
\includegraphics[width=5.in]{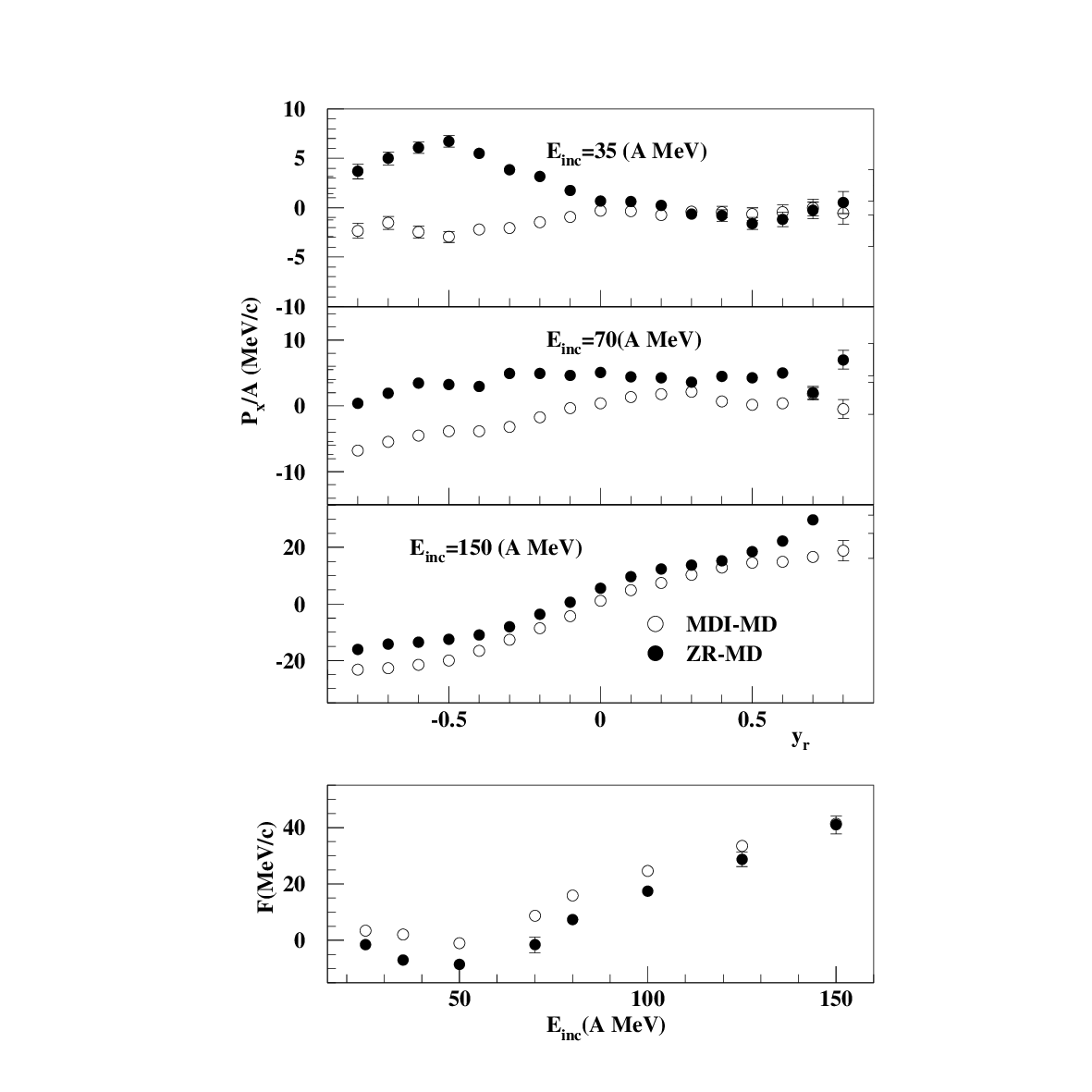}
\caption{\label{Fig:7}  In the three upper panels, for the zero an finite range interction (MDI)  different values of the nucleon effective mass the average transverse momentum per nucleon ($b=3$ fm)are plotted  as a function of the c.m. reduced rapidity.
In the bottom panel the slope parameters as  a function of the incident energy are also shown.
}
\end{figure}
\noindent

\section{Summary and final remarks}
Many-body correlations developed in the constrained Molecular Dynamics, have been analysed in the present work. 
The analysis proceeded by taking as reference a nuclear matter density functional  at zero temperature having commonly accepted properties around the saturation density.

\noindent
This correlations are evidenced by performing a comparison between the results obtained through the CoMD model using the same effective interaction as the one deduced in the MF approximation.

\noindent
In the case of  a finite range interaction or a MDI the comparison highlights large differences between the reference density functional  and the CoMD one. 
The latter in fact produces  different saturation density, binding energy, etc.
The sources of the correlations producing such differences have been  discussed. They arises from the wave-packed dynamics and from the constraint associated to the Pauli principle.

A procedure has been described to modify the values of the strength parameters in such
a way to obtain the reference EoS properties also in the  CoMD case.
The more simple case of a zero range interaction has been treated in Appendix B. In this case the changes
in the EoS are much more smaller and are due essentially due to the iso-vectorial interaction.

\noindent
Another part of the work was instead devoted to an attempt to understand the effects of these correlations on the dynamics of HIC at energies below 150 AMeV.
The studied system was
$^{64}Ni+^{48}Ca$  in central, semi-central collisions at different energies.
In particular a comparison between the case of zero and finite range interactions MDI-MD and ZR-MD (that are the interactions modified to take into account the CoMD correlations) has been performed  concerning  the predominant reaction mechanisms produced for central, semi-central collisions and for the production of nucleon transverse flow.
\noindent
The comparison highlights the repulsive character  of the studied finite range interaction in the expansion stage  with
respect the local one in producing lower fusion/incomplete-fusion associated to
a higher yield of IMF and to a higher fission probability.
Larger transverse flow is also produced in the finite range case.
These last calculations therefore clearly show that several common accessible observables in HIC display a rather high sensitivity to specific effects related to the range of the microscopic effective interaction. 
Moreover, for the investigated quantities, the study as function of the beam energy evidences  a typical 
energy beyond which these dependences are attenuated.
This behaviour could be  useful in trying to get information
about the effective interaction from the comparison between extended model calculations and experiments at different energies.
Concerning the many-body correlations, the comparison between the cases MDI-MD and MDI-MF (i.e. inclusion or not inclusion of the the corrections for CoMD correlations) shows  in the explored energy range a rather high sensitivity of the investigated measurable  quantities.
This sensitivity  is reduced for the cases ZR-MD and ZR-MF. 
\noindent
Finally, it's worth noting  that
even if the obtained numerical results are strictly valid for the specific model calculations,
the rather general feature of the discussed correlations could give a wider meaning to the 
relative changes obtained in the comparisons between the different studied cases.

\section{Appendix A}
In this section we discuss with some detail the Pauli constraint procedure in the box calculations.

\noindent
The occupation numbers $f_{i}$ in phase-space
around each particle are evaluated according to the \textit{method-2} of the CoMD description in Ref. \cite{tra2022}.
In particular the WP representative volume $V_{0}$ is an hyper-cube of
volume $h^{3}$
  with sides proportional to the WPs widths in phase space (same proportional constant in momentum and configuration space). 
The occupation is calculated as follows: $f_{i}=1+\dfrac{\sum^{'}_{j\ne i}\Delta V_{j,i}}{V_{0}}$. The apex indicates the sum on
identical particles. 
$\Delta V_{j,i}$ represents the generic overlap volume between the WPs $i$ and $j$.
When $f_{i}>f_{m}$ (see below) the constraint is activated.
As described in Refs.\cite{comd,miocomp} repeated 2-body elastic scatterings  are performed between the involved generic couples of particles $i$ and $k$ to lower the value of $f_{i}$.
The procedure is applied 10-20 times (depending on the density) and the configuration with the lower value of
 $f_{i}+f_{k}$ is accepted with final momenta lowe than the Fermi one.
Moreover, we have implemented the constraint by  adding another procedure to improve the Pauli prescription.
After the scattering procedure,  other better configurations are searched for:
the most near (in phase-space)identical particle $j$ to the particle $i$  exchange its charge or spin with
a close particle $k$ (in space). 
The particle $k$ has  to be distant (in phase-space) from the particle $i$.
The new configuration is accepted only if it produces a further lowering of 
$f_{i}+f_{k}$.
This procedure is attempted 5-10 times depending on the density.

\noindent
These transformations are applied in the PrS stage described in Sec. 2.4.
By considering different microscopic realizations of the system
and after the "stationary" conditions are reached, the lower value of $\overline{f}$ (the average value of the numbers $f_{i}$ for each realizzation) $f_{m}$, will be slightly larger than 1 ($\simeq 1.03$) as due to the finite efficiency and to the  numerical precision of the applied constraint procedures. 
Therefore, in general, if the so evaluated occupation numbers are used to decide whether allow or not a nucleon-nucleon collision, 
the blocking condition will be: $F_{i}=f_{i}-f_{m}>0$.
However, the routine that usually simulate  the collision integral is deactivated in this box calculations.
\noindent
We now present  some details concerning the box calculations and  the criteria used in searching for the minimum value of the total energy 
(within the numerical precision associated to the algorithm). 

\noindent
The box calculations have been performed by imposing Periodic
Boundary Conditions. For this calculations the routine that evaluate
the collision integral associated to the nucleon-nucleon collisions is  
This technique is widely used in numerical simulations of very large systems for which the effects related
to the surface are negligibly small also by using a limited
number of particles.
In our calculations we have distributed uniformly $A=2000$ nucleons (neutron and protons according
to $\beta$ parameter) into a cubic main cell having for a given density $\rho$ a side 
$L=(A/\rho)^{1/3}> 2\lambda$.
The momenta of the particles have been distributed according to
a Fermi distributions .
The equations of motion then have been solved according to the following condition on the Cartesian particle coordinates 
$c\equiv x,y,x$:
\begin{align}
if \hskip5pt c < -\dfrac{L}{2} \Longrightarrow c=c+L \\
if \hskip5pt c \geq \dfrac{L}{2} \Longrightarrow c=c-L  
\end{align} 
moreover, by indicating with $dc=c(j)-c(i)$ the generic relative coordinates and $|dc|\neq L$
\begin{align}
if \hskip5pt dc > \dfrac{L}{2} \Longrightarrow dc=dc-L \\
if \hskip5pt dc \leq -\dfrac{L}{2} \Longrightarrow dc=dc+L 
\end{align} 
In this way the interaction  with the particles of the 6 image cells
surrounding the main one are taken into account.

\noindent
After the PrS stage (see Sect. 2.4), in which the coordinates are not evolved in time but the constraint is rather  active to reduce the $\overline{f}$ value, the time evolution starts. 

\noindent
To search for the "ground state configuration" the time evolution is now coupled with a local cooling-warming procedure (particle by particle) to realize the condition
\begin{equation} 
{F}_{i}\approx 0
\end{equation}
for the largest number of particles. 

\noindent
In particular at each time step if $F_{i}>0$
(after the described constraint procedure), the momenta of the nucleons near to the nucleon $i$  in phase-space are scaled by a factor 1.01. If $F_{i}\leq 0$ the factor is 0.98 \cite{comd}.

\noindent
This procedure requires different time steps but it keeps the total energy and especially the average kinetic energy as low as possible in a consistent way with the Pauli principle.

As an example in Fig. 8 a) we show, as a function of time, the phase space occupation  $f_{i}$  for each neutron.
The calculations are referred to a reduced number of nucleons A=1000 ($\beta=0$) at the saturation density. The interaction in this case corresponds to the parameters shown in the second row of Table 1. 

\begin{figure}[htbp]
\includegraphics[width=5.1in]{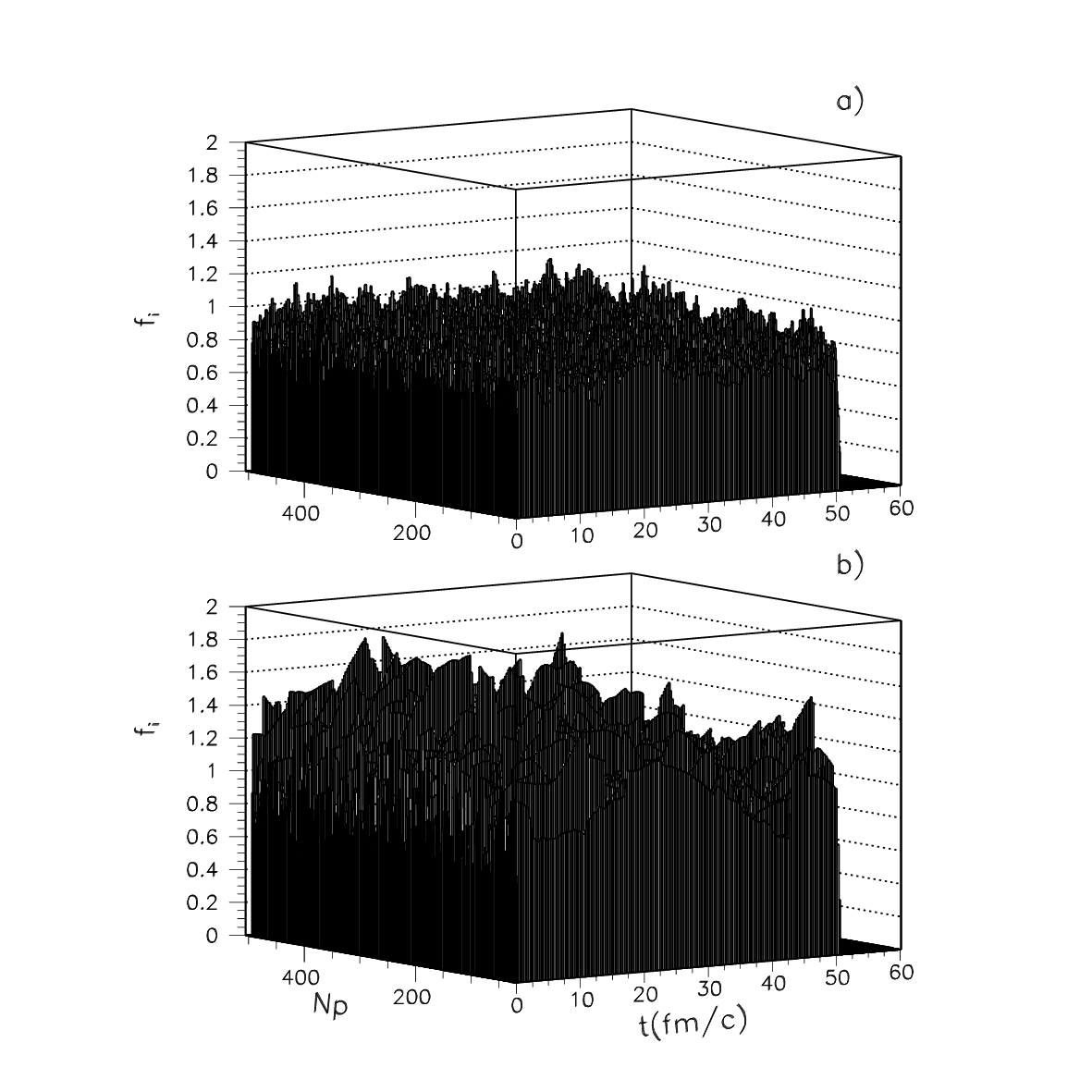}
\caption{\label{fig:A1} panel a):
By using the interaction parameters 
shown in the second row of Table 1  we plot the time dependence of the phase-space occupation $f_{i}$ for each particle (500 neutrons) at the saturation density.
panel b): same like panel a) but without the constraint} 
\end{figure}
\begin{figure}[htbp]
\includegraphics[width=5.1in]{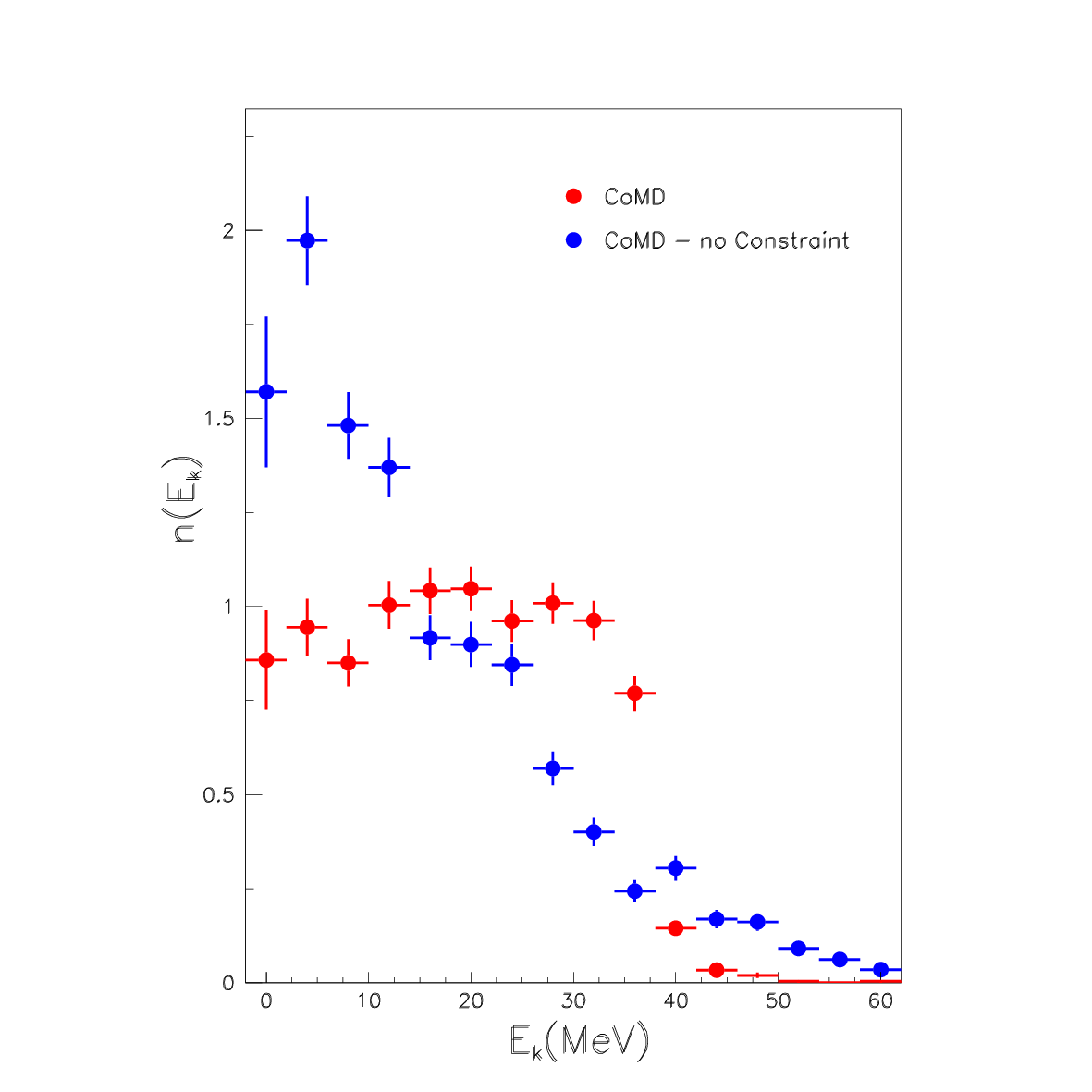}
\caption{\label{fig:A2}
 Nucleon occupation numbers $n(E_{k})$
as a function of the kinetic energy associated to 
the WP centroids for CoMD
calculations at t=50 fm/c  with and without the constraint. (color) } 
\end{figure}

\noindent
In Fig. 8 b) we instead show the same quantities without the constraint and we see clearly a violation of the Pauli prescription for many nucleons.
At 50 fm/c we show in Fig. 9 with red symbols
the nucleon occupation numbers as $n(E_{k})$
as a function of the kinetic energy associated to 
the WP centroids \cite{miocomp,bon2}.
In this case a Fermi-like distribution is obtained  with small deviations from a step-wise energy distributions which is instead  typical  of a non interacting Fermi-gas. 
A small depletion is seen at low energy compensated by
rather small values  beyond the Fermi energy.
These distortions are mainly produced through  the interaction between nucleons arising from fluctuations. The fluctuations  still exist in the local densities at the saturation density and can locally produce a gradients different from zero through which particles can exchange energy. Also this effect could be understood as a kind
of short range correlations.
The values
of the total energy obtained through this procedure, but using the parameters of
the first row in Table 1 are represented in
Fig. 1 with empty symbols.
Finally the blue markers in Fig. 9 represents the $n(E_{k})$ values obtained without the constraint. 
The distribution is now rather broad with an average kinetic energy larger than the previous case.
The distribution asymptotically can be associated to a large temperature in a Boltzmann statistical scheme \cite{miocomp,bon1} .

\noindent

\noindent
Therefore for densities greater than $\simeq 0.8\rho_{0}$, the obtained energy values from this procedure can be considered (within the limitation of the numerical procedure) as representative of the  minimum values ("ground states") associated to the Fermionic dynamics in a box.
Finally, the  microscopic configuration describing the particle distributions in phase-space are saved to be used in the next stage to find the new set of parameters as described in Sect. 2.5.
\section{Appendix B}
This section illustrates  the results obtained
by using a simple ZR interaction with 2 an 3-body terms
plus the iso-vectorial one. The reference EDF is the one producing an 
equilibrium density 
$\rho_{0}=0.165$ $fm^{-3}$, with an associated 
binding energy
$E(\rho_{0})=-16$ MeV,  compressibility $K(\rho_{0})=240$ MeV,  
 a symmetry energy $E_{sym}=30$ MeV, and a slope parameter
 $L=67.5$ MeV.
These features are the same that characterize the MDI-MD case whose
parameters fill the first row of Table 2.
The values of the strength parameters for the different terms obtained in the MF limit are: $P_{20}=-205.4$ MeV, $P_{30}=153.4$ MeV,
$P_{40}=36$ MeV, $\sigma=1.35$ and $\gamma=0.81$.

\noindent
On the panel a) of Fig. 10  the corresponding EDF as function of the relative density $\rho_{r}$  is plotted with a red line for symmetric NM.
Using the aforementioned  parameters in the CoMD box calculations and proceeding in the same way as the MDI
case (see also Sec. 2.4 and 2.5) we get the results
shown in the same panel with open symbols (ZR-MF).

\noindent
In the panel b) the related symmetry energy is shown.
The errors for each determination are within the symbol size. 
As discussed in Ref. \cite{miosky} these changes are mainly produced trough the constraint and the iso-vectorial interaction itself.  
 In our case it can be expressed as:
\begin{equation}
E_{sy} \approxeq \dfrac{P_{40}}{2\rho_{0}}
(\dfrac{\overline{S_{v}^{i}}}{\rho_{0}})^{\gamma-1}A\tilde{\rho}
[(1+\dfrac{\alpha}{2})\beta^{2}-\dfrac{\alpha}{2}]
\end{equation}
but $\overline{S_{v}^{i}}\approxeq \rho$ within 4\%.
So that with $\gamma\simeq 0.8$  we have:

\begin{equation}
E_{sy} \approxeq \dfrac{P_{40}}{2\rho_{0}}
(\dfrac{\rho}{\rho_{0}})^{\gamma-1}A\tilde{\rho}
[(1+\dfrac{\alpha}{2})\beta^{2}-\dfrac{\alpha}{2}]
\end{equation}
within 1\%.

\noindent
According to Eqs. (25-30) of Ref. \cite{miosky}  
$\tilde{\rho}$ is the average overlap integral per couples
of neutrons and protons.
$\alpha=\dfrac{\tilde{\rho}^{np}-\tilde{\rho}}{\tilde{\rho}}$
where $\tilde{\rho}^{np}$ is the average overlap per neutron-proton couples. 
The Pauli constraint and the iso-vectorial  interaction itself produce and $\alpha$ value  larger than zero.
This means that neutrons and protons  are slightly more overlapped with respect to the other couples ("deuteron effects"). Therefore, this produces
a change of the symmetry energy and a shift of the EDF for symmetric NM (second term in eq. (59)).
The obtained values of $\alpha$  around the saturation density are shown in the panel c) of Fig. 10.
For other densities shown in the figure the  values of $\alpha$ are obtained through an extrapolation based on a linear dependence from the relative density.
From Fig. 10 we can observe that for the ZR case the effects of the correlations
discussed in this  work are much smaller with respect to the MDI case.

\noindent
Subsequently we have applied the same procedure described in Sect. 2.6 to obtain the new set of values for the parameters to recover the
the reference EDF (red lines in Fig. 10 panel a) and b)).
The obtained results are plotted by means of black symbols (ZR-MD).
The new set of values is:
$P'_{20}=-202.4$ MeV, $P'_{30}=152.4$ MeV,
$P'_{40}=34.5$ MeV, $\sigma=1.38$ and $\gamma=0.86$.

\begin{figure}[htbp]
\includegraphics[width=5.in]{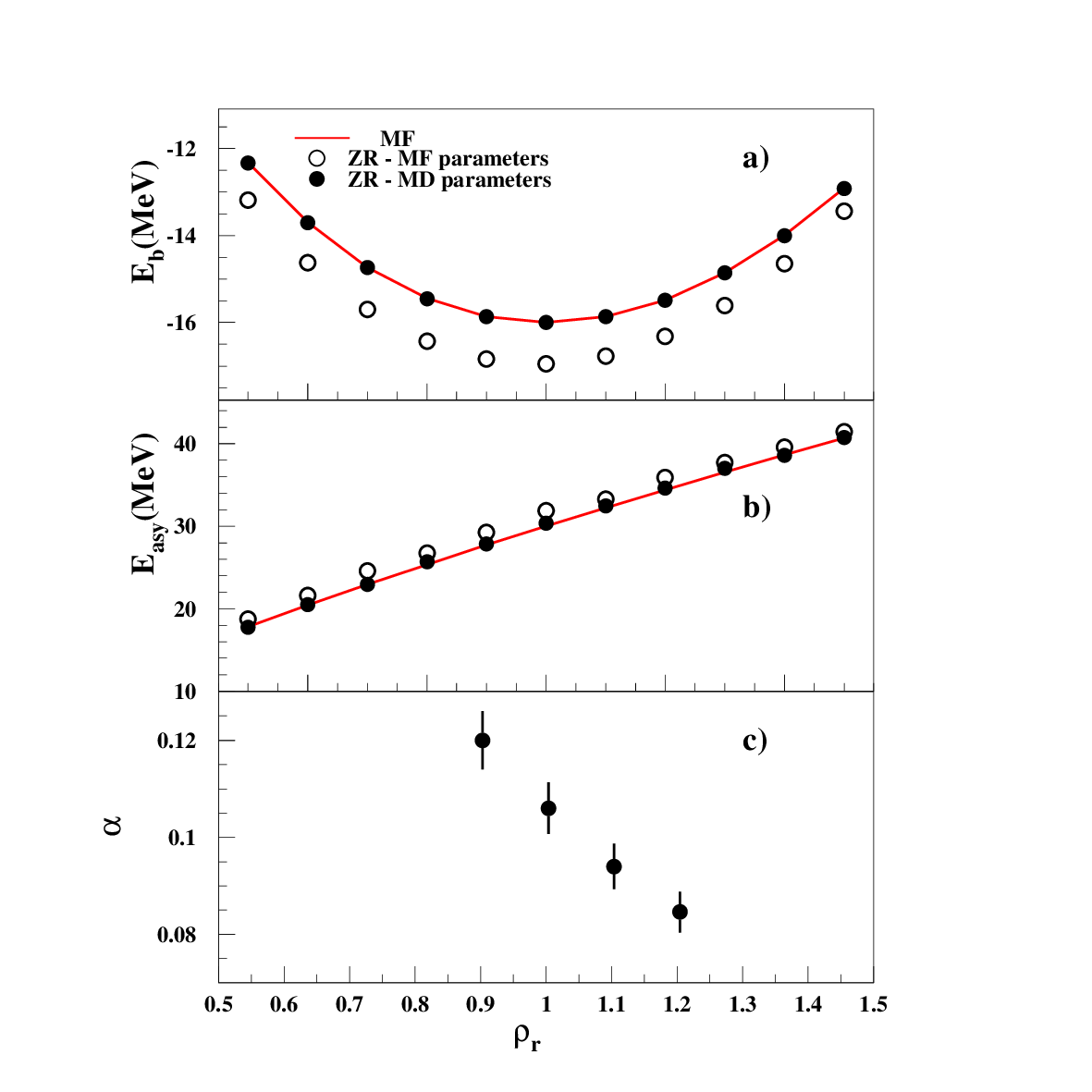}
\caption{\label{Fig:10} For symmetric NM, in the upper panel a) with a red line it is shown the total energy per nucleon $E_{b}$ as a function of the relative density $\rho_r$ evaluated in the MF approximation.  
 In the panel b) the corresponding values of the symmetry energy are also shown.
In the panel a) and b), 
the open symbols correspond to results of CoMD box calculations performed  using the parameter values obtained in the MF approximation (ZR-MF).
The solid represent the CoMD results obtained with the new set of strength parameter values reported in the section (ZR-MD). 
In the panel c) the values of the  $\alpha$
quantity are plotted for different relative densities (see text).  Errors are inside the size of symbols.
 (color)} 
\end{figure}

\noindent
Similarly to what discussed in Sect. 3, with these new parameters we have studied the $^{64}Ni+^{48}Ca$ collision at different incident energies. 
Good "ground state" configurations for the $^{64}Ni$ and $^{48}Ca$ nuclei have been obtained with a surface coefficient  $C_{s}=4.3$ MeVfm$^{2}$.
The calculations have been performed in the range of impact parameter
$b=0-5$ fm.
\noindent
In Fig. 11 as function of the incident energy  we plot the fusion cross section $\sigma_{f}$ and the fission probability $P_{f}$ for the ZR-MD case (blu dots) and for the 
ZR-MF case (red square).
Fig. 12 shows both the cross section  $\sigma_{IMF}$ and the multiplicity $M_{IMF}$ for the intermediate mass fragment production in the mid-rapidity region.
 Panel c) displays the ratio
$R=\dfrac{\sigma _{IMF}^{ZR-MD}}{\sigma _{IMF}^{ZR-MF}}$.

\begin{figure}[htbp]
\includegraphics[width=5.in]{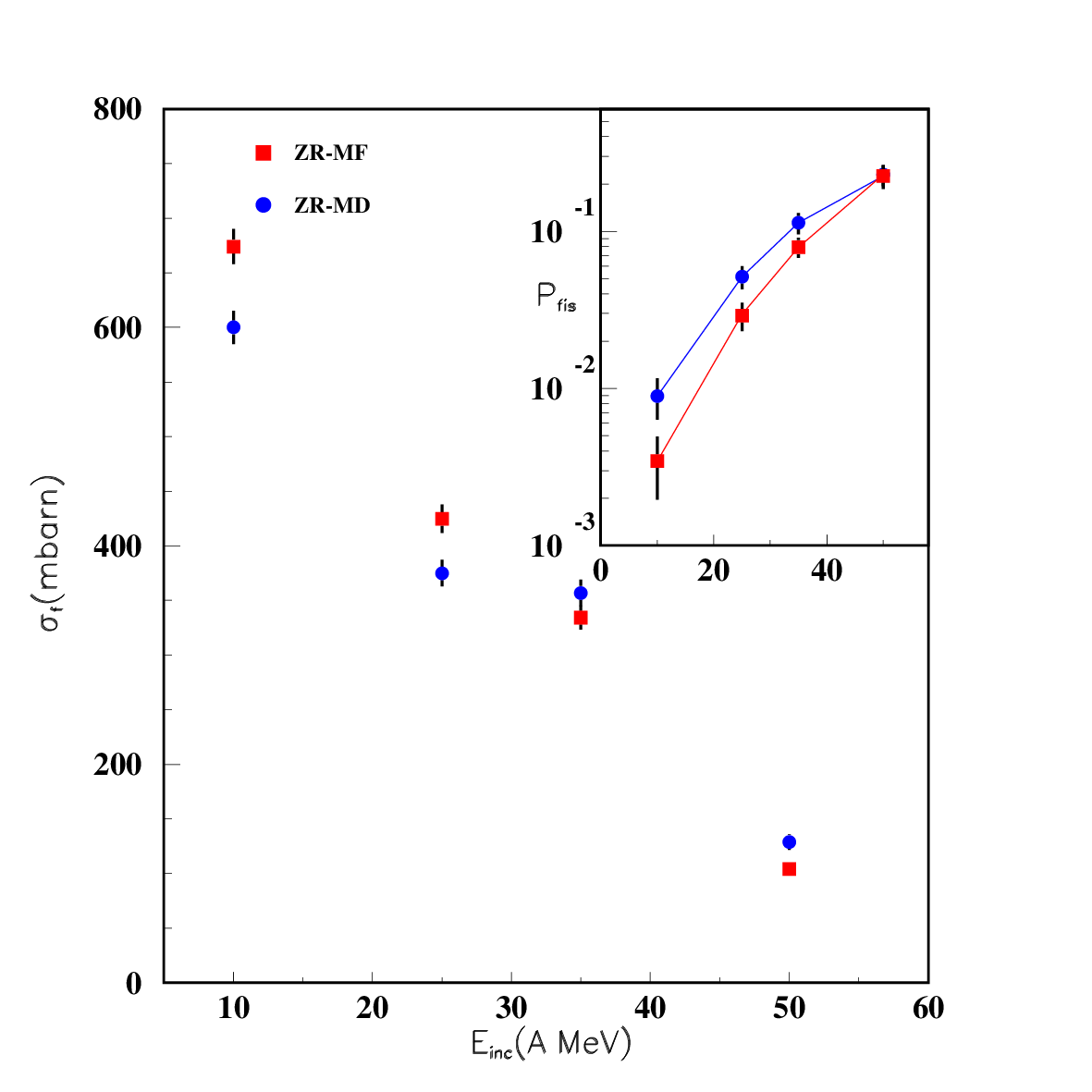}
\caption{\label{Fig:B1}  As a function of the incident energy $E_{inc}$ for the $^{64}Ni+^{48}Ca$ collision and
impact parameter range $b=0-5$ fm  the evaluated cross-section for the formation of an heavy residue is plotted.
The blue dots and the red square  symbols refer to the ZR-MD and ZR-MF cases respectively.
In the inset the fission probabilities are also shown.
The lines joining the points ZR-MD and ZR-MF are plotted to simplify the comparison between the two cases. (color on-line)}
\end{figure}
\noindent

\begin{figure}[htbp]
\includegraphics[width=5.in]{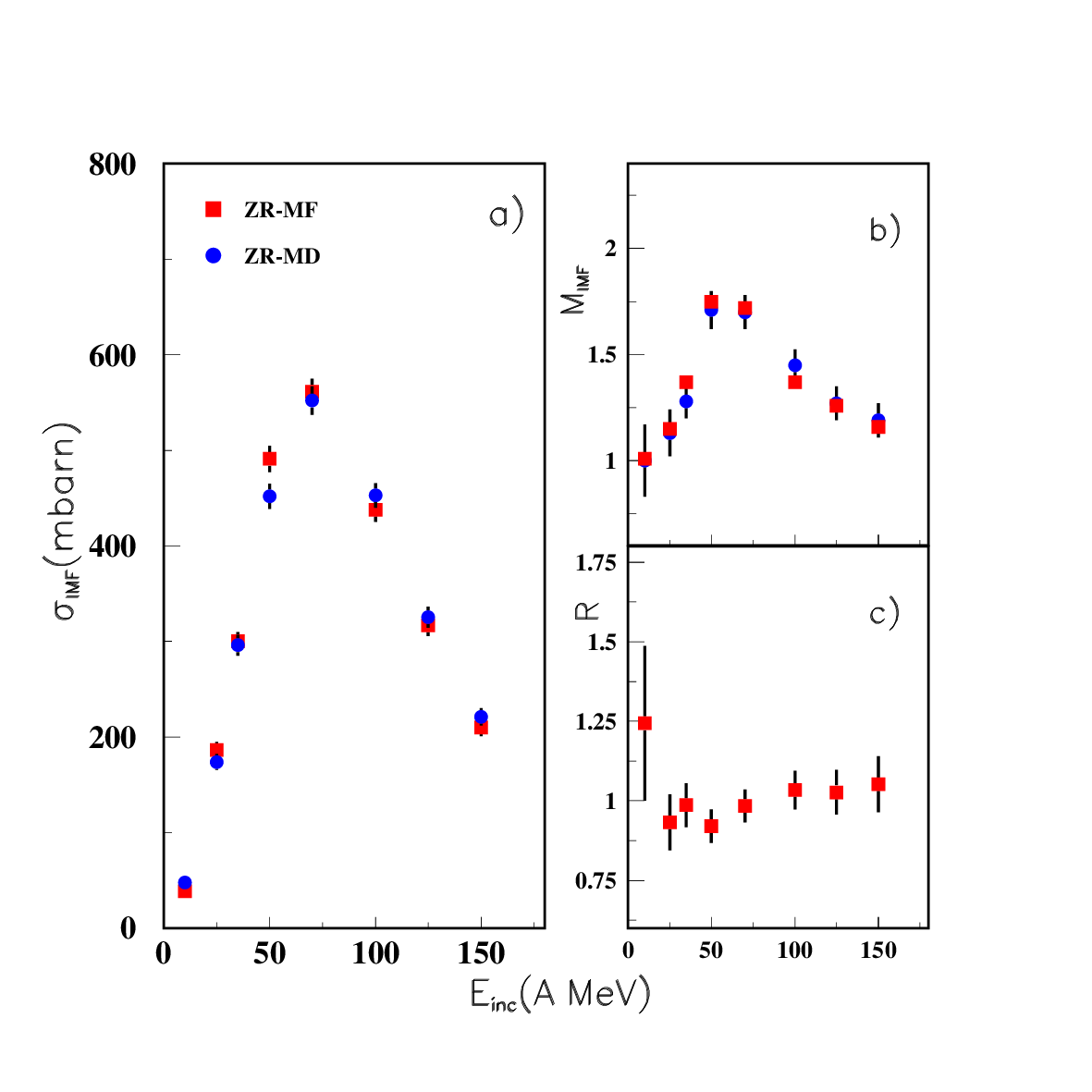}
\caption{\label{Fig:B2}  For the system $^{64}Ni+^{48}Ca$, in the impact parameters range  $b=0-5$ fm: panel a) shows for the ZR-MD, ZR-MF  cases the 
cross section for the production of at least one IMF  as a function of the incident energy $E_{b}$, panel b) shows the
associated multiplicity ZR-MD and ZR-MF , panel c) shows the ratio
$R=\dfrac{\sigma_{IMF}^{ZR-MD}}{\sigma_{f}^{ZR-MF}}$.
The lines joining the points ZR-MD and ZR-MF are plotted to highlight  the regular trend in the two cases.(color on -line)}
\end{figure}

\noindent
As it can be expected from the NM calculations the differences induced on these measurable quantities are rather small with respect to the MDI cases especially concerning the IMF production which seems to be almost insensitive to the change of parameters.
Differences due to the discussed correlations are instead still present
in the formation of hot-source in the mid-rapidity region at a level of 10-25\% in the interval $E_{inc}$ 10-50 AMeV.
The associated  fission probabilities show a larger sensitivity.

\section{Appendix C}

\begin{figure}[htbp]
\includegraphics[width=5.1in]{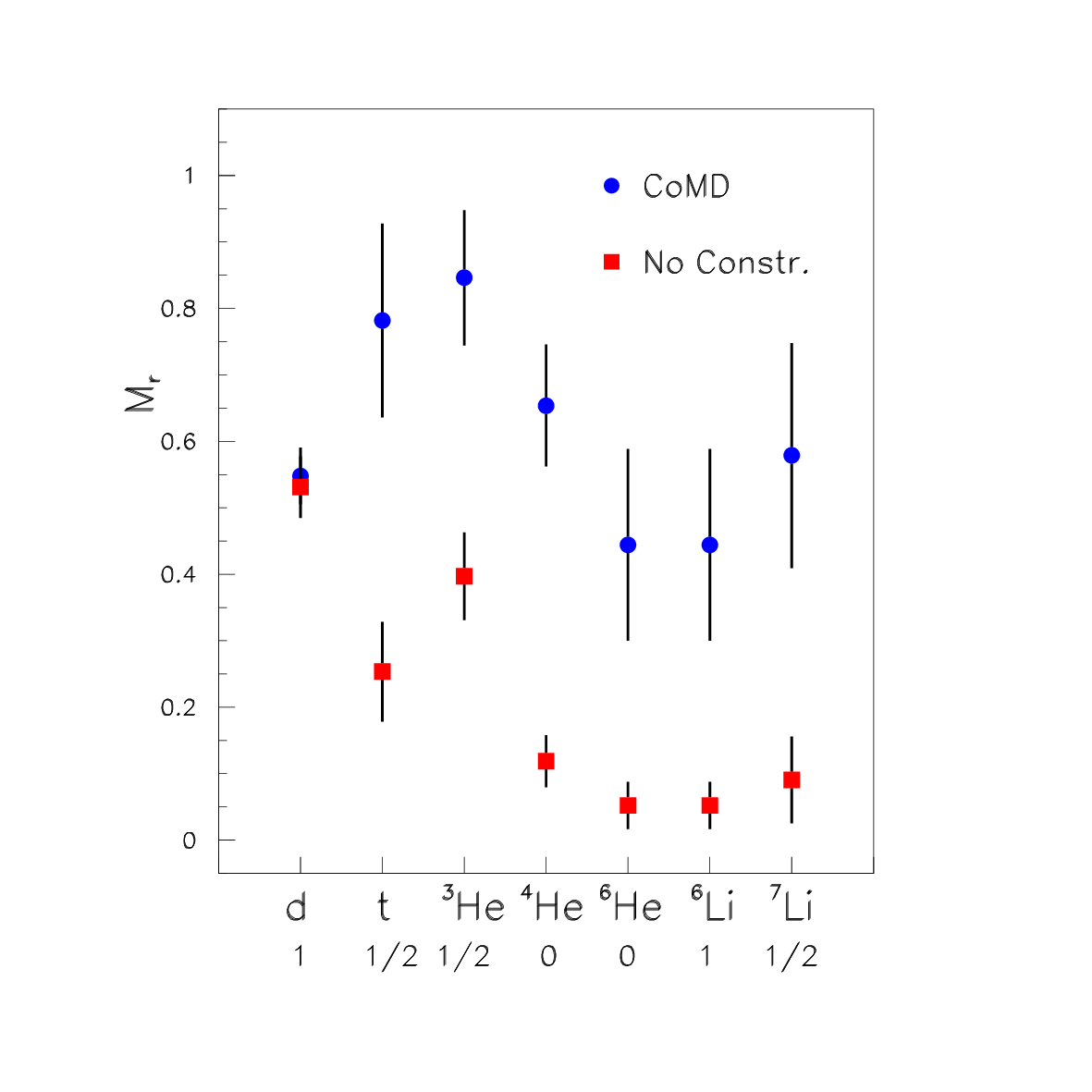}
\caption{\label{fig:B1} The relative produced yields $M_{r}$ are shown for fully CoMD  and no constraint calculations at $\rho=0.2\rho_{0}$.
For each isotope, whose chemical symbol is plotted along the horizontal axes, $M_{r}$ represents the ratio between the yield of isotope having the right spin and the related total yield. The value of the right spin is reported under each chemical symbol. (color
on-line)} 
\end{figure}
This section illustrates the effects produced by the constraint
based on the Pauli principle on the  light cluster production taking into account their intrinsic spin, at low density.
To this aim we performed box CoMD calculations for five independent configurations  with and without the constraint at a density equal to 0.2$\rho_{0}$. 
We let the system evolve for 20 fm/c, after that we have searched 
for cluster production by applying the Minimum Spanning Tree method
every 5 fm/c upto 45 fm/c. We have used  a coalescence radius equal to 2.1 fm.
For each light isotope produced we have evaluated
the fraction $M_{r}$ of fragments having the right total spin with respect to the total number of the produced isotope having any possible
combination of spin. 
These quantities averaged on  5 time steps and on the different independent realizations are shown in Fig. 10 for the both cases.
As can be seen fully CoMD calculations largely enhance the rate $M_{r}$
with respect to the the no-constraint case. This effect is directly related
to the enhanced probability to obtain in  primary clusters doublets and quartets of neutrons or protons  with zero total spin which are near in phase-space.

\end{document}